\documentclass[aps,reprint]{revtex4-2}
\usepackage[utf8]{inputenc}
\usepackage{amssymb}
\usepackage{amsmath}
\usepackage{graphicx}
\usepackage[caption=false,font=small,subrefformat=parens,labelformat=parens]{subfig}
\usepackage[colorlinks=true,linkcolor=blue,allcolors=blue]{hyperref}
\usepackage{nicefrac}
\usepackage{float}
\usepackage{comment}
\newcommand\numberthis{\addtocounter{equation}{1}\tag{\theequation}}
\usepackage{physics}
\usepackage{gensymb}
\newcommand{\mbfk}{\mathbf{k}}
\usepackage{siunitx}

\usepackage{pifont}
\usepackage{bbding}

\usepackage{subfiles}

\def\biblio{\bibliographystyle{apsrev4-2}\bibliography{biblio}}

\begin{document}
\def\biblio{}

\title{Valley-polarization in biased bilayer graphene using circularly polarized light}
\author{A. Friedlan}
\email{alex.friedlan@queensu.ca}
\author{M. M. Dignam}
\affiliation{Department of Physics, Engineering Physics \& Astronomy, Queen’s University, Kingston, Ontario, Canada, K7L 3N6}

\begin{abstract}
    Achieving a population imbalance between the two inequivalent valleys is a critical first step for any valleytronic device. A valley-polarization can be induced in biased bilayer graphene using circularly polarized light. In this paper, we present a detailed theoretical study of valley-polarization in biased bilayer graphene. We show that a nearly perfect valley-polarization can be achieved with the proper choices of external bias and pulse frequency. We find that the optimal pulse frequency $\omega$ is given by $\hbar\omega=2a,$ where $2a$ is the potential energy difference between the graphene layers. We also find that the valley-polarization originates not from the Dirac points themselves, but rather from a ring of states surrounding each. Intervalley scattering is found to greatly reduce the valley-polarization for high frequency pulses. Thermal populations are found to significantly reduce the valley-polarization for small biases. This work provides insight into the origin of valley-polarization in bilayer graphene and will aid experimentalists seeking to study valley-polarization in the lab.
\end{abstract}
\maketitle

\section{Introduction}\label{Intro}
Since its first realization in 2004 \cite{graphene2004}, graphene has promised to revolutionize electronics with its high electron mobility \cite{Mobility,rozhkov}, impressive mechanical strength \cite{Strong}, and tunable Fermi level \cite{CastroReview}. There exist two inequivalent local minima in graphene's band structure known as \mbox{\textit{valleys}} or \textit{Dirac points} which we label $K$ and $K'.$ In analogy with spintronics \cite{Spintronics}, the valley index is binary and the concept of using this two-state system to perform logical operations is known as valleytronics \cite{small}. To realize such a system, we require a way to induce a valley-polarization, that is, a differential electron population between the $K$ and $K'$ valleys. 

There have been many proposals for valleytronic devices based on monolayer graphene, however most have relied on configurations that may be difficult to realize in the lab \cite{golub2011valley,Rycerz,Barriers,Strain,Teardownthatwall}. Intrinsically, the $K$ and $K'$ valleys are indistinguishable from one another. This means that it is difficult to selectively populate the valleys, say, using an optical field. Inversion symmetry breaking is necessary for graphene-based valleytronics \cite{yao2007valley,yao2008valley}. One solution is to use a staggered sublattice potential, for instance, by growing graphene on a substrate of hBN \cite{staggered}. Another option is to consider materials with intrinsically broken inversion symmetry. TMDs such as monolayer MoS\textsubscript{2} have gained significant interest recently, in part due to the presence of an intrinsic band gap at the Dirac points \cite{caovalleyselective,makcontrol,zengMoS2,mak2014}. In this work we consider bilayer graphene, which consists of two graphene sheets stacked in an AB/Bernal stacking arrangement \cite{hopping}. Biasing the bilayer by applying a potential difference across the two graphene sheets breaks the inversion symmetry and opens a band gap \cite{falcopunch,Asymmetry,gateinduced,directobs}. Not only that, but the band gap can be tuned continuously from zero to the mid-infrared by adjusting the strength of the external bias \cite{controlling,EFE,makobs}. For an excellent review of the electronic properties of both monolayer and bilayer graphene, please see McCann \cite{McCannTB}.

It has been proposed that circularly polarized light can be used preferentially inject carriers into the $K$ and $K'$ valleys of bilayer graphene \cite{yao2008valley}. Right-hand circularly polarized light couples strongly to the $K$ valley, while light of the opposite helicity couples strongly to $K'.$ There has been significant work towards inducing valley-polarized currents in bilayer graphene with broken inversion symmetry \cite{LiValve,Shimazaki,suigate}, but very few studies have focused on using circularly polarized light to induce a valley-polarization \cite{abergel,abergel2011}. To the best of our knowledge, no studies have yet sought to maximize the optically-induced valley-polarization, leaving experimentalists ill-equipped to study this phenomenon in the laboratory. Several important questions remain unanswered: What is the optimal operating external bias? What is the optimal operating pulse frequency? And what pulse duration should be used? It is also to date unknown as to which scattering processes fundamentally limit performance of bilayer-graphene-based valleytronic devices: How clean a sample is required? Can a valley-polarization be observed at room temperature? In this paper, we seek to answer these questions as well as offer valuable insight into the underlying physics of valley-polarization in bilayer graphene.

Our findings can be summarized as follows. At low temperatures, and in the absence of scattering, a near-perfect valley-polarization can be obtained for pulse frequencies $\omega$ satisfying $\hbar\omega=2a,$ where $2a$ is the potential energy difference between the graphene layers. This result originates from a $k$-dependent valley-contrasting optical selection rule that becomes exact when $\hbar\omega=2a.$ This finding is qualitatively consistent with some previous calculations \cite{yao2008valley,abergel}, but so far seems to have gone unnoticed in the literature. Our calculations indicate that intervalley scattering via optical phonons greatly reduces the valley-polarization when operating at high pulse frequencies. We also find that thermal electron populations significantly reduce the valley-polarization for small external biases. Taken together, intervalley scattering and thermal electron populations complicate the simple picture that the valley-polarization is maximized along $\hbar\omega=2a.$ In all cases, to maximize the valley-polarization, the pulse duration should be close to or larger than the sample decoherence time. For typical samples, with the proper choice of pulse frequency and external bias, a valley-polarization of up to 70\% can be achieved at room temperature. At low temperatures ($<150$ K), the valley-polarization can be as large as 97\%. 

This paper is organized as follows. In Sec. \ref{Theory}, we present our theoretical model. We construct the graphene Hamiltonian and solve for the energy bands and eigenstates. We then develop our density matrix equations of motion, solving them perturbatively for excitation by circularly polarized light. In Sec. \ref{Results}, we study the resulting valley-polarization, primarily as a function of the frequency of the exciting field and of the external bias between the graphene layers. We first examine a simplified model before proceeding to introduce intervalley scattering and thermal effects. We also examine the effects of varying the pulse duration and decoherence time before concluding in Sec. \ref{Conclusion}.

\section{Theory}\label{Theory}
We employ a four-band nearest-neighbor tight-binding model to calculate the low-energy electron bands and Bloch eigenstates. We perturb the system with an optical field, treating the interaction within the length gauge. We develop density matrix equations of motion and solve them up to second-order. We calculate the electron populations in the $K$ and $K'$ valleys that result from the linear absorption of a circularly polarized Gaussian pulse.

\subsection{Tight-binding}
\begin{figure}
\includegraphics[width=\linewidth]{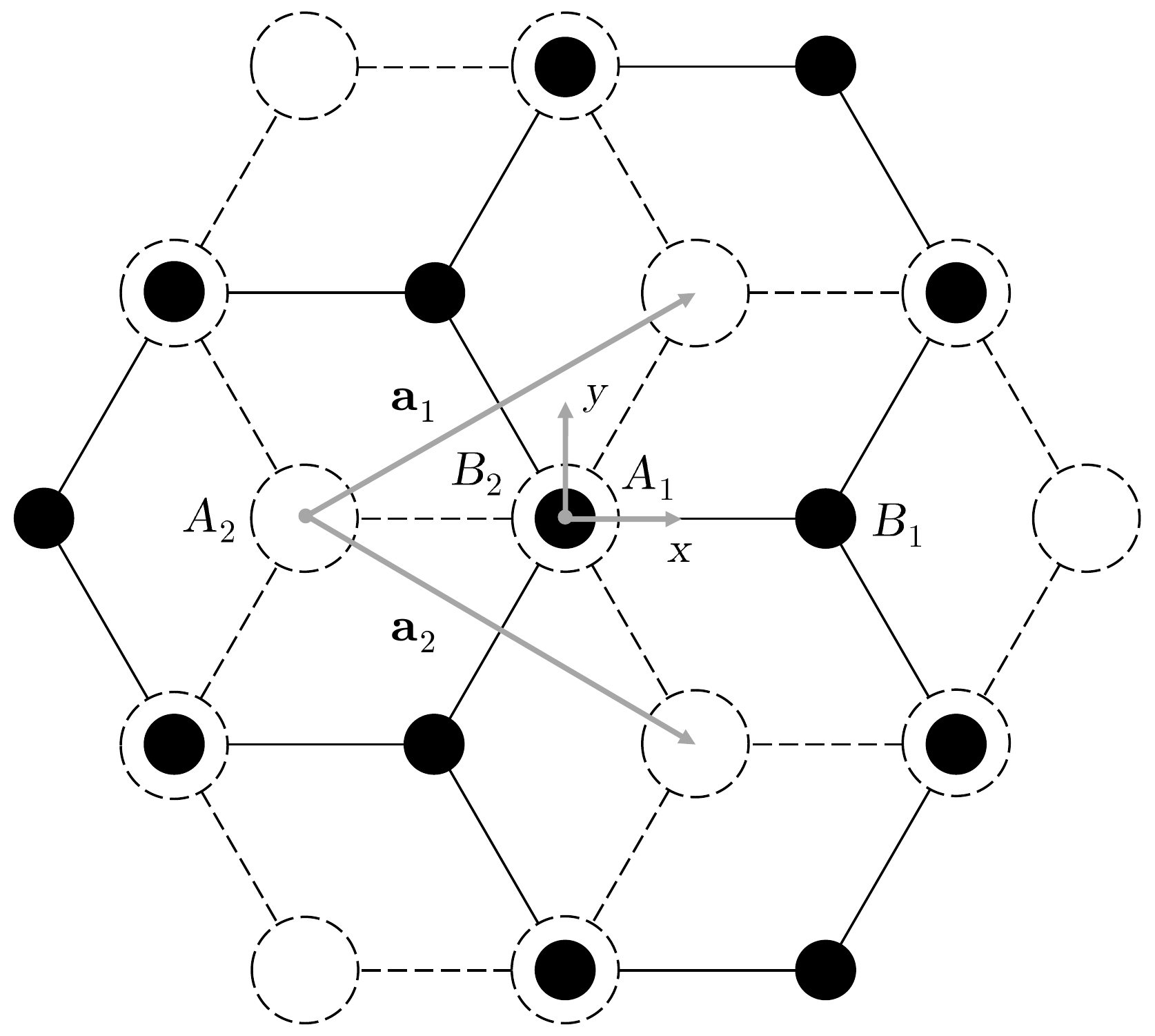}
\centering
\caption{Bilayer graphene lattice. The $A_1$ and $B_1$ atoms are in the top layer at energy $+a$ (black), while the $A_2$ and $B_2$ atoms are in the bottom layer at energy $-a$ (white).}
\label{NamingConvention}
\end{figure}

We use as our basis the single-atom Bloch functions
\begin{equation}
    \Phi_i(\mathbf{k},\mathbf{r})=\sum_{j=1}^Ne^{i\mathbf{k}\cdot\mathbf{R}_j}\phi(\mathbf{r}-\mathbf{R}_{j,i}),
\end{equation}
where $\mbfk$ is the Bloch wave vector, $\mathbf{r}$ is the position vector, and $\phi(\mathbf{r})$ is a carbon $2p_z$ orbital  \cite{McCannTB}. The sum is over the $N$ different unit cells, and $\mathbf{R}_{j,i}\equiv \mathbf{R}_j+\mathbf{r}_i,$ where $\mathbf{R}_j$ is a Bravais lattice vector and $\mathbf{r}_i$ is a basis vector, which denotes the position of one of the four atoms in the unit cell. Following the coordinate conventions of Ref. \cite{riley2016}, we have ${\mathbf{r}_{A_1}=d\mathbf{\hat{z}}},$ $\mathbf{r}_{B_1}=a_0\mathbf{\hat{x}}+d\mathbf{\hat{z}},$ $\mathbf{r}_{A_2}=-a_0\mathbf{\hat{x}}-d\mathbf{\hat{z}},$ and ${\mathbf{r}_{B_2}=-d\mathbf{\hat{z}},}$ where $a_0=1.42$ \AA\, is the interatomic distance, and $2d$ is the interlayer spacing (see Fig. \ref{NamingConvention}).

Using these basis states, we construct our eigenstates
\begin{equation}\label{Bloch}
    \Psi_{n\mathbf{k}}(\mathbf{r})=\braket{\mathbf{r}}{n\mbfk}=A_n(\mathbf{k})\sum_{i}C_{n}^i(\mathbf{k})\Phi_i(\mathbf{k,\mathbf{r}}),
\end{equation}
where $A_n(\mathbf{k})$ is a normalization factor, the $C_{n}^i(\mbfk)$ are expansion coefficients, $i$ indexes the atoms, and $n$ labels the band. In the basis $i=\{A_1,B_1,A_2,B_2\},$ including hopping between nearest-neighbors within each layer, and between the overlapping $A_1$ and $B_2$ atoms in opposite layers, we obtain the nearest-neighbor tight-binding Hamiltonian  \cite{McCannTB}
\begin{equation}
    \mathcal{H}_{0}=\left(\begin{array}{cccc}
a & f(\mathbf{k}) t_{\|} & 0 & t_{\perp} \\
f^*(\mathbf{k}) t_{\|} & a & 0 & 0 \\
0 & 0 & -a & f(\mathbf{k}) t_{\|} \\
t_{\perp} & 0 & f^*(\mathbf{k}) t_{\|} & -a
\end{array}\right),
\end{equation}
where (by convention) $2a\geq 0$ is the potential energy difference between the graphene layers, and where $t_{\|}=3.3$ eV and $t_\perp=0.42$ eV are, respectively, the intra- and inter-layer hopping energies \cite{hopping}. The function ${f(\mbfk)=1+e^{-i\mbfk\cdot\mathbf{a}_1}+e^{-i\mbfk\cdot\mathbf{a}_2}}$ describes hopping between nearest-neighbor sites, where ${\mathbf{a}_1=a_0(3\mathbf{\hat{x}}+\sqrt{3}\mathbf{\hat{y}})/2}$ and ${\mathbf{a}_2=a_0(3\mathbf{\hat{x}}-\sqrt{3}\mathbf{\hat{y}})/2}$ are the primitive translation vectors (see Fig. \ref{NamingConvention}). In what follows, we will focus on the dynamics in the vicinities of the Dirac points ${\mathbf{K}= 4\pi\hat{\mathbf{y}}/3\sqrt{3}a_0}$ and ${\mathbf{K'}=- 4\pi\hat{\mathbf{y}}/3\sqrt{3}a_0}.$ To obtain the Hamiltonian for electrons close to these points, we expand $f(\mbfk)$ about $\mathbf{K}$ and $\mathbf{K'}$ and obtain ${f(\mathbf{k})\approx i\frac{3}{2}a_0ke^{\pm i\theta_k},}$ with the plus and minus signs corresponding to the $K$ and $K'$ valleys respectively. Note that we have transformed to a polar coordinate system with origin at $\mathbf{K}$ or $\mathbf{K'},$ where $k=|\mbfk|=(k_x^2+k_y^2)^{1/2},$ and $\theta_k$ is the angle $\mbfk$ makes with the $k_x$ axis. The function $f(\mbfk)$ may also be expressed in terms of the graphene Fermi velocity $v_f=3a_0t_\|/2\hbar\approx 10^6$ \si{\meter\per\second} according to $f(\mbfk)=i\hbar v_fke^{\pm i\theta_k}/t_\|.$

\begin{figure}
\includegraphics[width=\linewidth]{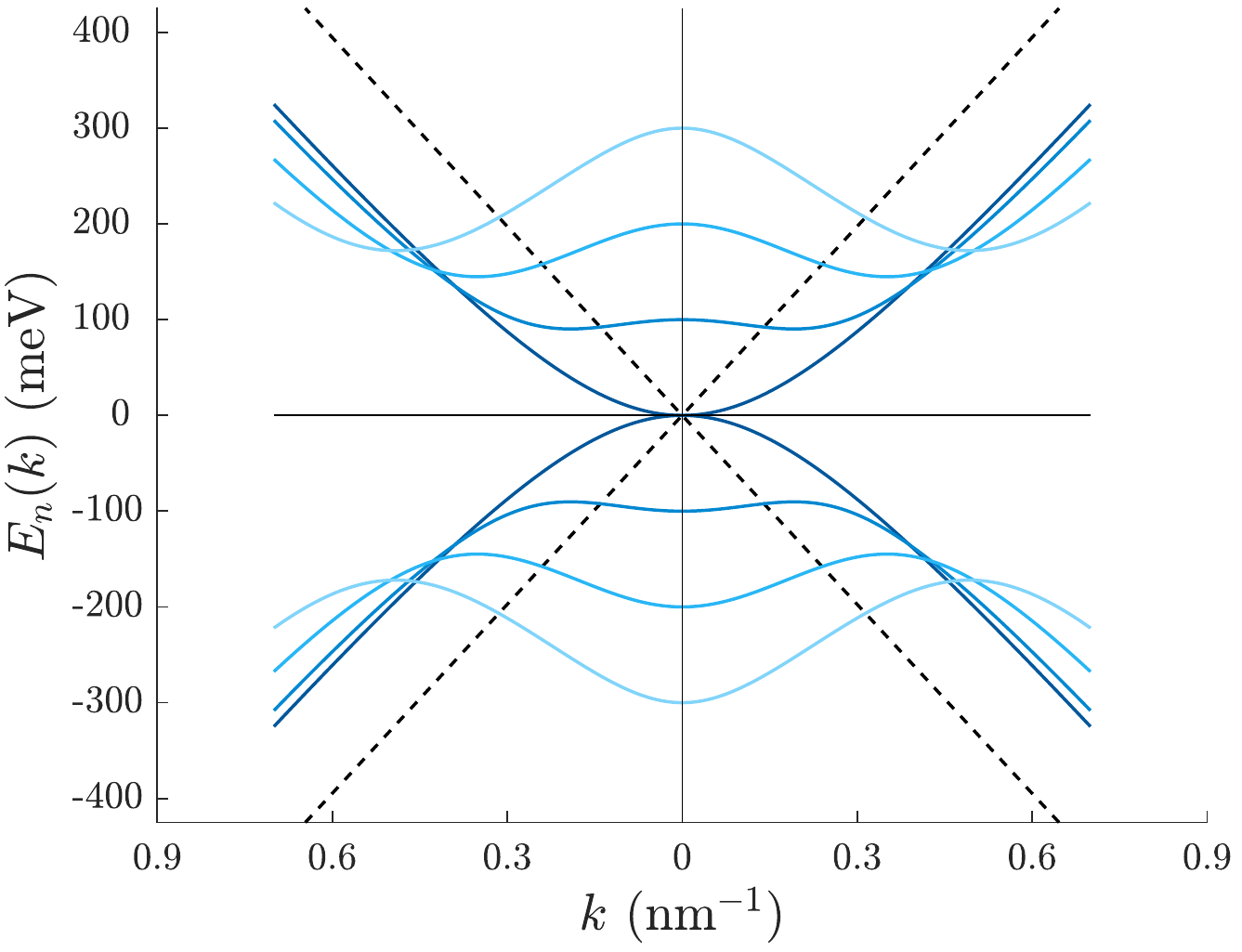}
\centering
\caption[Bilayer Graphene Basis Definitions]{The conduction and valence band energies as a function of $k$ for four different biases $a.$ From darkest to lightest, $a=0, 100, 200$ and $300$ meV. For comparison, the intersecting dashed lines are the energy bands of monolayer graphene.}
\label{BandStructure}
\end{figure}

Neglecting overlap between inequivalent atoms, we solve for the energies $E_n(\mbfk)$ and eigenvectors $\Psi_{n\mbfk}(\mathbf{r})$ of $\mathcal{H}_0.$ The dynamics of the system will be studied using only the two lowest-energy bands whose (dimensionless) energies are given by \cite{Asymmetry}
\begin{equation}
    \widetilde{E}_{n}(\mbfk)=\widetilde{E}_{n}(k)=\pm \sqrt{\tilde{t}^2|f(\mathbf{k})|^2+\tilde{a}^2+\frac{1}{2}-\tilde{\epsilon}(\mbfk)},
\end{equation}
where $\widetilde{E}_n(k)\equiv E_n(k)/t_\perp,$ where $n=\{c,v\}$ labels the conduction band and valence band, and where we have introduced the dimensionless quantities ${\tilde{a}\equiv a/t_\perp},$ ${\tilde{t}\equiv t_{\|}/t_\perp,}$ and ${{\tilde{\epsilon}(\mbfk)\equiv((4\tilde{a}^2+1)\tilde{t}^2|f(\mathbf{k})|^2+1/4)^{1/2}.}}$ The conduction and valence bands are shown in Fig. \ref{BandStructure} for four different biases.
The energy bands are isotropic in-plane, but we show the bands reflected across the $k=0$ axis to emphasize the symmetry. The dispersion is electron-hole symmetric. At the Dirac points ($k=0)$, $E_c(k)=a$ and $E_v(k)=-a.$ The band gap for a particular bias $a$ is given by \cite{McCannTB}
\begin{equation}\label{bandgap}
    \Delta E(a) =\frac{2at_\perp}{\sqrt{4a^2+t_\perp^2}}.
\end{equation}
The expansion coefficients of $\Psi_{n\mbfk}(\mathbf{r})$ are
\begin{align*}
     C_{n}^{A_1}(\mathbf{k}) & =\widetilde{E}_{n}(\mbfk)-\tilde{a}, \\
    C_{n}^{B_1}(\mathbf{k}) & = \tilde{t}f^*(\mathbf{k}), \\
    C_{n}^{A_2}(\mathbf{k}) & = \frac{C_{n}^{B_2}(\mathbf{k)}   \big(C^{B_1}_{n}(\mathbf{k})\big)^*}{\widetilde{E}_{n}(\mbfk)+\tilde{a}}, \\ C_{n}^{B_2}(\mathbf{k}) & = \big(C_{n}^{A_1}(\mathbf{k})\big)^2-\big|C_{n}^{B_1}(\mathbf{k})\big|^2.
    \numberthis{}
\end{align*}

\subsection{Connection elements}\label{Connection}
We treat the carrier-field interaction using the length gauge Hamiltonian $\mathcal{H}_L=-e\mathbf{E}(t)\cdot\mathbf{r},$ where $e=-|e|$ is the electron charge, $\mathbf{E}(t)$ is the (classical) electric field of the optical pulse at the graphene, and $\mathbf{r}$ is the electron position operator. The density matrix equations of motion we will derive in the following section require matrix elements of $\mathbf{r}.$ Following Blount \cite{Blount}, we have
\begin{equation}\label{rmatele}
    \langle n\mathbf{k}|\mathbf{r}|m\mathbf{k'}\rangle =i\delta_{nm}\nabla_{\mathbf{k}}\delta(\mathbf{k-k'})+\delta(\mathbf{k-k'})\boldsymbol{\xi}_{nm}(\mathbf{k}),
\end{equation}
where we have defined the connection elements
\begin{equation}\label{xinm}
    \boldsymbol{\xi}_{nm}(\mathbf{k})\equiv i\frac{(2\pi)^2}{\Omega}\int d^3\mathbf{r}\, u_{n\mathbf{k}}^*(\mathbf{r})\nabla_{\mathbf{k}}u_{m\mathbf{k}}(\mathbf{r}),
\end{equation}
where $\Omega$ is the area of a real-space unit cell, and the integration is over $\Omega,$ and over $-\infty<z<\infty$ perpendicular to the plane. The cell-periodic function $u_{n\mathbf{k}}(\mathbf{r})$ is defined by
\begin{equation}\label{unk}
    \Psi_{n\mathbf{k}}(\mathbf{r})=e^{i\mathbf{k}\cdot\mathbf{r}}u_{n\mathbf{k}}(\mathbf{r}).
\end{equation}
Imposing orthonormality on $\Psi_{n\mbfk}(\mathbf{r})$, we obtain
\begin{equation}
    \int d^3\mathbf{r}\,u_{n\mbfk}^*(\mathbf{r})u_{m\mbfk}(\mathbf{r})=\frac{\Omega}{(2\pi)^2}\delta_{nm},
\end{equation}
and, in the nearest-neighbor tight-binding approximation, the normalization factor
\begin{equation}
    A_n(\mbfk)=\frac{\sqrt{\Omega}}{2\pi}\left(\sum_i \big|C_{n}^i(\mbfk)\big|^2\right)^{-1/2}.
\end{equation}

\begin{figure}
\includegraphics[width=\linewidth]{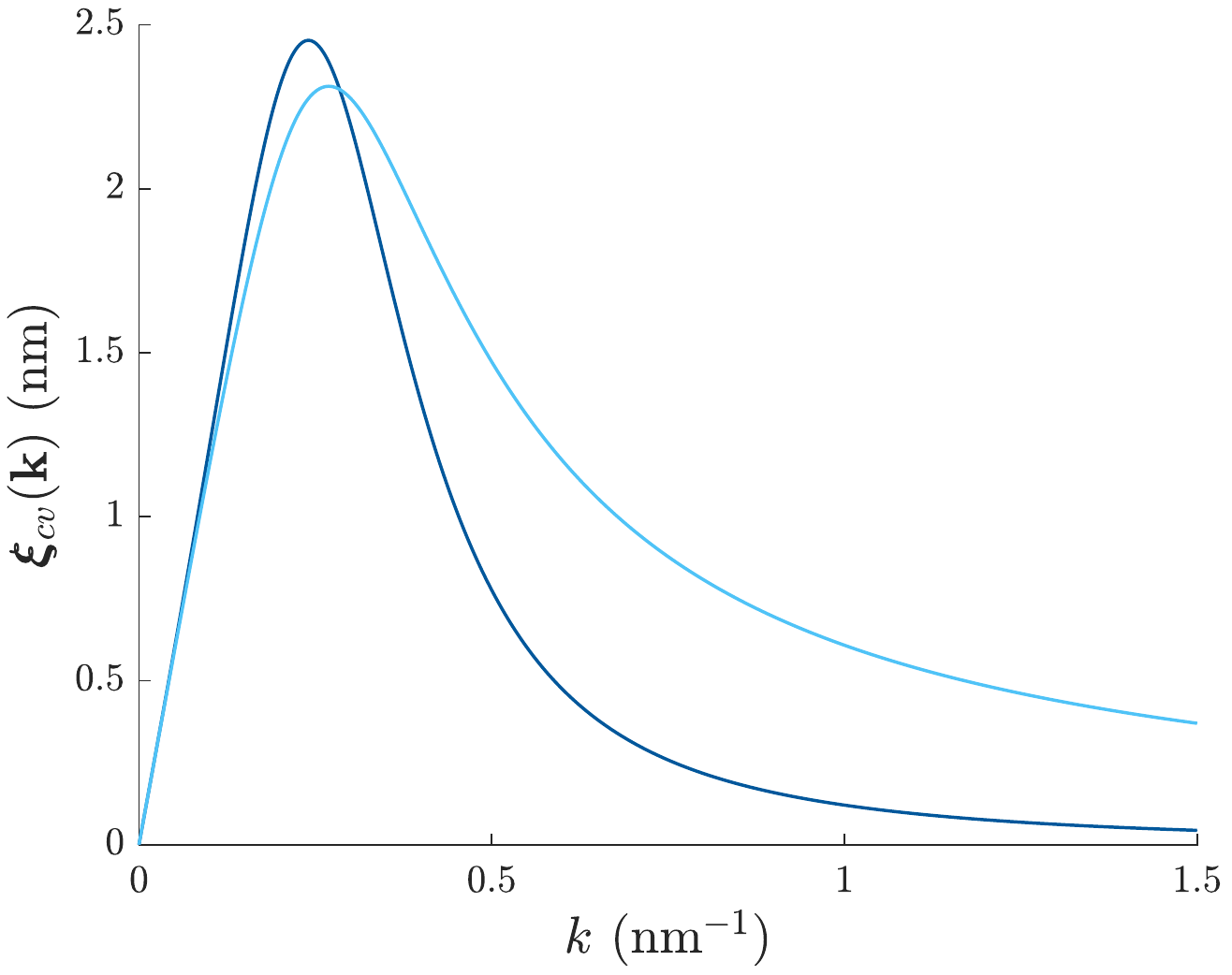}
\centering
\caption[AB Functions]{The connection element functions $A$ (dark) and $B$ (light) as a function of $k$ for $a=100$ meV.}
\label{ABFuncs}
\end{figure}

The connection element for transitions between the conduction band and valence band is \footnote{Intraband connection elements, i.e. $\boldsymbol{\xi}_{nn}(\mbfk)$, will have an extra term within the summation proportional to the gradient of the normalization factor $A_n(\mathbf{k})$}
\begin{multline}
\label{xivc}
\boldsymbol{\xi}_{cv}(\mathbf{k})=i\frac{(2\pi)^2}{\Omega}A_c^*(\mathbf{k})A_v(\mathbf{k})\\\times\sum_{i}\big(C^{i*}_{c}(\mathbf{k})\nabla_\mathbf{k}C_{v}^i(\mathbf{k})-i\mathbf{r}_iC^{i*}_{c}(\mathbf{k})C_{v}^i(\mathbf{k})\big).
\end{multline}
In what follows, we neglect the term proportional to $\mathbf{r}_i$ in Eq. (\ref{xivc}), as we neglect terms of similar magnitude when we expand $f(\mathbf{k})$ to first-order in $k$ about the Dirac points. Eq. (\ref{xivc}) can be evaluated analytically, but as the expression is very long we do not present it here explicitly. We find that when $f(\mbfk)$ is taken to first-order, the connection element takes the form
\begin{equation}\label{xiAB}
    \boldsymbol{\xi}_{cv}(\mbfk)=iA(a,k)\mathbf{\hat{k}}\pm B(a,k)\boldsymbol{\hat{\theta}}_k,
\end{equation}
where $\mathbf{\hat{k}}$ and $\boldsymbol{\hat{\theta}}_k$ are the standard unit vectors in our polar coordinate system, and where the plus and minus signs correspond to the $K$ and $K'$ valleys respectively. Here, $A$ and $B$ are real, positive functions that depend only on the magnitude of $\mathbf{k}$ and the external bias $a.$ Although $\boldsymbol{\xi}_{cv}(\mbfk)$ depends on the choice of gauge (i.e. the $\mbfk$-dependent phase of the Bloch eigenstates $\ket{n\mbfk}$), the action of a gauge transformation is to simply multiply $\boldsymbol{\xi}_{cv}(\mbfk)$ by an overall phase factor \cite{Blount}, rendering the magnitudes of $A$ and $B,$ as well as the ratio $A/B,$ gauge-invariant quantities. We plot $A$ and $B$ as functions of $k$ in Fig. \ref{ABFuncs} for an example bias of ${a=100}$~meV.
As can be seen, $A=B=0$ at $k=0$ (i.e. at the Dirac points) \footnote{$B=0$ at $k=0$ only when $a$ is nonzero. If $a=0,$ $B\rightarrow\infty$ at $k=0$.}. Besides $k=0,$ there is second $k$ for which $A=B,$ whose value depends on the external bias. In the large-$k$ limit, $B\sim1/k,$ just as it does in monolayer graphene \cite{HighHarmonic}. Biased bilayer graphene is unlike monolayer graphene or unbiased bilayer graphene ($a=0$) in that $A$ is nonzero \cite{HighHarmonic,riley2016,rileypaper}. In the limit of large $k,$ $A\sim1/k^3.$ Under $a\rightarrow -a,$ $A$ is unchanged but $B\rightarrow-B.$ We will return these functions in Sec. \ref{Results} when we discuss the results of our calculations. As we shall see, it is the non-vanishing of $A$, and the crucial sign difference between the $K$ and $K'$ valleys that unlocks the possibility of valley-polarization.

\subsection{Equations of motion}
The Heisenberg equations of motion for the reduced density operator $\rho$ in the basis of Bloch states $\ket{n\mbfk}$ in the relaxation time approximation are \cite{HighHarmonic,AversaSipe95}
\begin{align*}\label{rhonm}
    &\frac{d\rho_{nm}(\mathbf{k})}{dt}=-i\omega_{nm}(\mathbf{k})\rho_{nm}(\mbfk)-\frac{e}{\hbar}\mathbf{E}(t)\cdot\nabla_\mbfk\rho_{nm}(\mbfk)\\& \qquad\qquad i\frac{e}{\hbar}\mathbf{E}(t)\cdot\sum_l\left[\boldsymbol{\xi}_{nl}(\mathbf{k})\rho_{lm}(\mathbf{k})-\boldsymbol{\xi}_{lm}(\mathbf{k})\rho_{nl}(\mathbf{k})\right]\\&\qquad\qquad\qquad\qquad-\gamma_{nm}(\mbfk)\left[\rho_{nm}(\mathbf{k})-\rho_{nm}^{eq}(\mbfk)\right],\numberthis{}
\end{align*}
where $\hbar\omega_{nm}(\mbfk)\equiv E_n(\mbfk)-E_m(\mbfk),$ $\rho_{nm}^{eq}(\mbfk)=\rho_{nn}^{eq}(\mbfk)\delta_{nm}$ is the equilibrium density matrix, and $\gamma_{nm}(\mbfk)$ is a matrix of scattering rates. If $n=m,$ we refer to $\gamma_{nm}(\mbfk)$ as the intraband scattering rate. If $n\neq m,$ we refer to $\gamma_{nm}(\mbfk)$ as the interband decoherence rate. We allow for the $\gamma_{nm}(\mbfk)$ to be in principle $k$-dependent but will save our discussion of these quantities for Sec. \ref{Results}. Using first-order perturbation theory, we solve Eq. (\ref{rhonm}) subject to the initial conditions
\begin{align*}
\begin{cases}
    \rho_{nn}^{eq}(\mbfk)=f_n(\mbfk) & \smash{\raisebox{-1.6ex}{at $t=-\infty,$}} \\
    \rho_{nm}^{(1)}(\mbfk)=0 \numberthis{}
\end{cases}
\end{align*}
where $f_n(\mbfk)=f_n(k)$ is the Fermi-Dirac distribution. To zeroth-order we obtain $\rho_{nm}^{(0)}(\mbfk)=f_n(\mbfk)\delta_{nm}.$ Using standard techniques, the first-order result for the off-diagonal density matrix element at time $t$ is found to be
\begin{align*}\label{rhocv}
    &\rho_{cv}^{(1)}(\mbfk)=e^{-\left(i\omega_{cv}(\mbfk)+\gamma_{cv}(\mbfk)\right)t}\int_{-\infty}^ti\frac{e}{\hbar}\mathbf{E}(t')\\&\qquad\cdot\boldsymbol{\xi}_{cv}(\mbfk)\big[f_v(\mbfk)-f_c(\mbfk)\big]e^{\left(i\omega_{cv}(\mbfk)+\gamma_{cv}(\mbfk)\right)t'}dt'.\numberthis{}
\end{align*}
We are interested in the \textit{total} electron populations around the $K$ and $K'$ valleys, not the $k$-space distributions. Therefore, when calculating the populations to second-order, we neglect the second term in Eq. (\ref{rhonm}) because this term simply redistributes carrier momentum within each valley. We find that the second-order contribution to the conduction band population is
\begin{align*}\label{rhocc}
    &\rho_{cc}^{(2)}(\mbfk)=e^{-\gamma_{cc}(\mbfk) t}\int_{-\infty}^t i\frac{e}{\hbar}\mathbf{E}(t')\\&\qquad\cdot\big[\boldsymbol{\xi}_{cv}(\mbfk)\rho_{vc}^{(1)}(\mbfk)-\boldsymbol{\xi}_{vc}(\mbfk)\rho_{cv}^{(1)}(\mbfk)\big]e^{\gamma_{cc}(\mbfk) t'}dt',\numberthis{}
\end{align*}
where $\boldsymbol{\xi}_{vc}(\mbfk)=\boldsymbol{\xi}_{cv}^*(\mbfk).$ Although $\boldsymbol{\xi}_{nm}(\mbfk)$ is gauge-dependent, $\rho^{(1)}_{mn}(\mbfk)$ transforms in the opposite sense and so the product $\boldsymbol{\xi}_{nm}(\mbfk)\rho^{(1)}_{mn}(\mbfk)$ is gauge-invariant. The $m$\textsuperscript{th}-order contribution to the carrier density in the conduction band about each Dirac point is given by
\begin{equation}\label{nvgen}
    n_V^{(m)}(t)=\frac{2}{(2\pi)^2}\int \rho_{cc}^{(m)}(\mbfk)\,d^2\mbfk,
\end{equation}
where $V=\{K,K'\}$ labels the valley, the factor of two accounts for spin degeneracy, and the integration is only in the vicinity of the particular Dirac point. The zeroth-order contribution gives the thermal carrier density, while the second-order contribution gives the injected carrier density.

\subsection{Calculating the injected carrier density}\label{Calculating}
To evaluate the injected carrier density, we must specify the electric field of the optical pulse. It is well known that a valley-polarization can be induced in biased bilayer graphene using circularly polarized light \cite{yao2008valley}. In Appendix \ref{Elliptical}, we discuss the possibility of using light of a more general elliptical polarization, but here we limit our discussion to circularly polarized light as we find circularly polarized light to always be optimal. We take $\mathbf{E}(t)$ to be a right-hand circularly polarized Gaussian pulse with central frequency $\omega,$ amplitude $E_0,$ and pulse duration $t_p.$ Thus, 
\begin{equation}
    \mathbf{E}(t)=E(t)\big(\mathbf{\hat{x}}+i\mathbf{\hat{y}}\big)e^{-i\omega t} + c.c.,
\end{equation}
where $E(t)=E_0e^{-t^2/t_p^2}.$ Given the form of the connection element [Eq. (\ref{xiAB})], it will be convenient to express the field in the local $k$-space coordinate system. Taking the origin to be either $\mathbf{K}$ or $\mathbf{K}',$
\begin{equation}\label{field}
    \mathbf{E}(t)=E(t)\big(\mathbf{\hat{k}}+i\boldsymbol{\hat{\theta}}_k\big)e^{-i\left(\omega t-\theta_k\right)}+c.c.
\end{equation}

\noindent With Eqs. (\ref{rhocc}) and (\ref{nvgen}), one has, for the injected carrier density,
\begin{align*}
     &n_V^{(2)}(t)=\frac{2}{(2\pi)^2}\int_{0}^{\infty}k\,dk\int_{0}^{2\pi}d\theta_k \,e^{-\gamma_{cc}(\mbfk) t}\int_{-\infty}^{t}dt^{\prime}\numberthis{}\\&\qquad\times i\frac{e}{\hbar}\mathbf{E}(t^{\prime})\cdot \big[\boldsymbol{\xi}_{cv}(\mathbf{k})\rho_{vc}^{(1)}(\mbfk)-\boldsymbol{\xi}_{vc}(\mathbf{k})\rho_{cv}^{(1)}(\mbfk) \big]e^{\gamma_{cc}(\mbfk) t^{\prime}}.
\end{align*}
Since the energy bands are isotropic, we assume that the scattering rates are as well so that we may write $\gamma_{nm}(\mbfk)=\gamma_{nm}(k).$ We may then pull $e^{-\gamma_{cc}(\mbfk)t}$ through the angular integral, and interchange the order of the temporal and angular integration. Performing the angular integral first, one can show (after significant work)
\begin{widetext}
\begin{align*}\label{erf1}
    n_V^{(2)}(t)&=\frac{2}{(2\pi)^2}\pi^{3/2}\frac{e^2}{\hbar^2}E_0^2t_p\int_{0}^{\infty}k\,dk\,e^{-\gamma_{cc} t} \int_{-\infty}^{t}dt^{\prime}\\&\qquad\times\left[\left(A\pm B\right)^2\left(\textrm{erf}\left(\frac{t^{\prime}}{t_p}+\lambda_-\right)+1\right)\textrm{exp}\left(\lambda^2_-+\frac{t^{\prime}}{t_p}\delta_--\frac{t^{\prime \,2}}{t_p^2} \right)\right.\\&\qquad\qquad\left.
    +\left(A\mp B\right)^2\left(\textrm{erf}\left(\frac{t^{\prime}}{t_p}+\lambda_+\right)+1\right)\textrm{exp}\left(\lambda^2_++\frac{t^{\prime}}{t_p}\delta_+-\frac{t^{\prime \,2}}{t_p^2}\right)\right]+ c.c.,\numberthis{}
\end{align*}
\end{widetext}
where we have suppressed explicit $k$-dependence, and where the upper and lower operations correspond to the $K$ and $K'$ valleys respectively. The Greek letters represent the quantities
\begin{align}\label{greek}
    \lambda_\mp & \equiv -\frac{t_p}{2}\left(\gamma_{cv}+i(\omega_{cv}\mp\omega)\right),\\
    \delta_\mp & \equiv 2\lambda_\mp+\gamma_{cc}t_p,
\end{align}
where the subscripts indicate whether the term is resonant ($-$) or anti-resonant ($+$) with the optical field. 

We are faced with the following integral:
\begin{align*}\label{precompl}
    J(t)&=\int_{-\infty}^{t}\,dt^{\prime}\left(\textrm{erf}\left(\frac{t^{\prime}}{t_p}+\lambda\right)+1\right)\\&\qquad\qquad\quad\times\textrm{exp}{\left(\lambda^2+\frac{t^{\prime}}{t_p}\delta-\frac{t^{\prime \,2}}{t_p^2}\right)}.\numberthis{}
\end{align*}
Making the replacement $\eta=t'/t_p+\lambda,$
and completing the square in the exponential, Eq. (\ref{precompl}) may be re-expressed as
\begin{align*}\label{postcompl}
    J(t)&=t_p\,\textrm{exp}(\beta)\int_{-\infty+\lambda}^{{t/t_p+\lambda}}\,d\eta\left(\textrm{erf}\left(\eta\right)+1\right)\\&\qquad\qquad\qquad\qquad\qquad\times\textrm{exp}\left(-(\eta+c)^2\right),\numberthis
\end{align*}
where
\begin{align*}\label{candbeta}
    \beta & \equiv  \lambda^2+\frac{1}{4}\delta^2, \\
    c & \equiv-\lambda-\frac{1}{2}\delta. \numberthis{}
\end{align*}
To the best of our knowledge, the integral in Eq. (\ref{postcompl}) cannot be performed analytically. However, an analytic result is provided by Ref. \cite{erftable} in the limit $t\rightarrow\infty.$ In Appendix \ref{CSI}, we address the subtleties of taking this limit given that $\lambda$ is in general complex. We find
\begin{equation}\label{erftoerfc}
    J(t\rightarrow\infty)=\sqrt{\pi}\,t_p\,\textrm{exp}(\beta)\,\textrm{erfc}\left(\frac{c}{\sqrt{2}}\right).
\end{equation}
For times $t=t_f$ by which the integrand of Eq. (\ref{postcompl}) has decayed essentially to zero, we may make the approximation $J(t_f)\approx J(t\rightarrow\infty).$ This approximation limits the applicability of this analytic result to times $t_f$ that are at least several times the pulse duration $t_p.$ This should not pose a problem in practice, because one would likely only wish to manipulate the valley-polarized carriers after the exciting pulse has passed. The pulse durations we will be considering are only on the order of 10s to a few 100s of femtoseconds, so this delay is very small. Using Eq. (\ref{erftoerfc}) in Eq.~(\ref{erf1}), we obtain the key result of this section: the carrier density injected into valley $V$ at time $t=t_f,$
\begin{widetext}
\begin{equation}\label{nsubV}
    n_V^{(2)}=\frac{1}{2}\frac{e^2}{\hbar^2}E_0^2\,t_p^2\int_0^\infty k\,dk\,e^{-\gamma_{cc} t_f}\left[\left(A\pm B\right)^2\text{exp}\,(\beta_-)\,\text{erfc}\left(\frac{c_-}{\sqrt{2}}\right)+\left(A\mp B\right)^2\text{exp}\,(\beta_+)\,\text{erfc}\left(\frac{c_+}{\sqrt{2}}\right)\right]+c.c.\numberthis{}
\end{equation}
\end{widetext}
Again, the upper and lower operations correspond to the $K$ and $K'$ valleys respectively, and the subscripts indicate whether the term is resonant ($-$) or anti-resonant ($+$) with the field. Recalling that $A$ and $B$ are real and positive, observe that in the $K$ valley, the resonant term couples to the large $(A+B)^2$ term, while the anti-resonant term couples to the small $(A-B)^2$ term. The situation is reversed in $K',$ in that it is the anti-resonant term which couples to $(A+B)^2,$ while the resonant term couples to $(A-B)^2.$ This asymmetry leads to a stronger response in $K$ than in $K',$ or in other words, a valley-polarization. For unbiased bilayer graphene ($a=0$), Eq. (\ref{nsubV}) should return equal populations for the $K$ and $K'$ valleys. Indeed, if $a=0,$ then $A(a,k)=0$ and the asymmetry between $K$ and $K'$ disappears. As a check for our code, we have confirmed that Eq. (\ref{nsubV}) returns equal populations in $K$ and $K'$ for the $a=0$ case when integrated numerically.

\section{Results}\label{Results}
We wish to compare the carrier densities in the conduction bands of the $K$ and $K'$ valleys shortly after excitation by a pulse of circularly polarized light. Depending on the temperature, a significant contribution to the carrier density can come from the zeroth-order thermal population $n_V^{(0)}.$ We will consider this in more detail in Sec. \ref{Density}, but for now we focus solely on the second-order response. To this end, we define the second-order valley-polarization $\mathcal{P}^{(2)}$ to be the difference between the carrier densities injected around the $K$ and $K'$ valleys at time $t=t_f$, normalized by their sum:
\begin{equation}\label{P2}
    \mathcal{P}^{(2)}\equiv\frac{n^{(2)}_K-n^{(2)}_{K'}}{n^{(2)}_K+n^{(2)}_{K'}}.
\end{equation}
When the system is completely valley-polarized in favor of $K$ ($K'$) electrons, $\mathcal{P}^{(2)}=1$ ($-1$). If the system is not valley-polarized, $\mathcal{P}^{(2)}=0.$ Throughout this section, we consider a (Gaussian) right-hand circularly polarized pulse, so we expect the system to be valley-polarized in favor of $K$ electrons.

We vary the external bias and pulse frequency to examine how the valley-polarization depends on these two parameters. We consider the frequency-bias pairs that result in the strongest valley-polarization to be the optimal operating parameters for valleytronic devices. Unless otherwise stated, we take the temperature $T$ to be $300$ K and the chemical potential to be at the charge-neutrality point ($\mu=0$). However, we emphasize that the second-order valley-polarization is only weakly dependent on $T$ and $\mu.$ Because of electron-hole symmetry, the choice $\mu=0$ leads to identical results for both electron and hole populations. For this reason, we discuss only electron populations. In what follows, we restrict ourselves to external biases greater than $50$ meV, because for lower biases, thermal populations and spatial variations in the system gating can severely limit the valley-polarization.

The relaxation dynamics of photoexcited carriers in monolayer graphene have been studied extensively \cite{Andreas,snapshots,Direct,Tracking}. For bilayer graphene, the relaxation processes are expected to be similar, but the associated time scales are not precisely known and depend on the substrate, fabrication process, and bias. The general consensus in the literature is that an initial period of rapid thermalization occurs due primarily to electron-electron scattering in which the photoexcited carriers relax to a quasi-thermal equilibrium within a few 10s of femtoseconds \cite{hof,inversion,giant}. This process is then followed by a period of carrier cooling due to the emission of optical phonons over a timescale of 100s of femtoseconds. Finally, there is a period of cooling due to acoustic phonon scattering and carrier recombination on the scale of picoseconds. In this work, because we are interested in the valley-polarization a few 10s to 100s of femtoseconds after the pulse arrives, we neglect these picosecond processes completely and focus on the early-time dynamics. As we are working in the perturbative regime, we will consider only intraband scattering processes.

To help develop the main ideas, in Sec. \ref{NoScattering} we work with a simplified model in which we neglect intraband scattering, accounting only for interband decoherence. In Sec. \ref{IntrabandScattering}, we will introduce intraband scattering via optical phonons. In particular, we will allow for intervalley scattering, which acts to reduce the valley-polarization, and for \textit{intra}valley scattering, which limits the \mbox{intervalley} process. In Sec. \ref{Effects}, we will examine the effects of varying the pulse duration and decoherence time. In Sec.~\ref{Density}, we will examine the effects of the thermal electron populations on the valley-polarization.

\subsection{Without intraband scattering}\label{NoScattering}
\begin{figure}
    \centering
    \includegraphics[width=\linewidth]{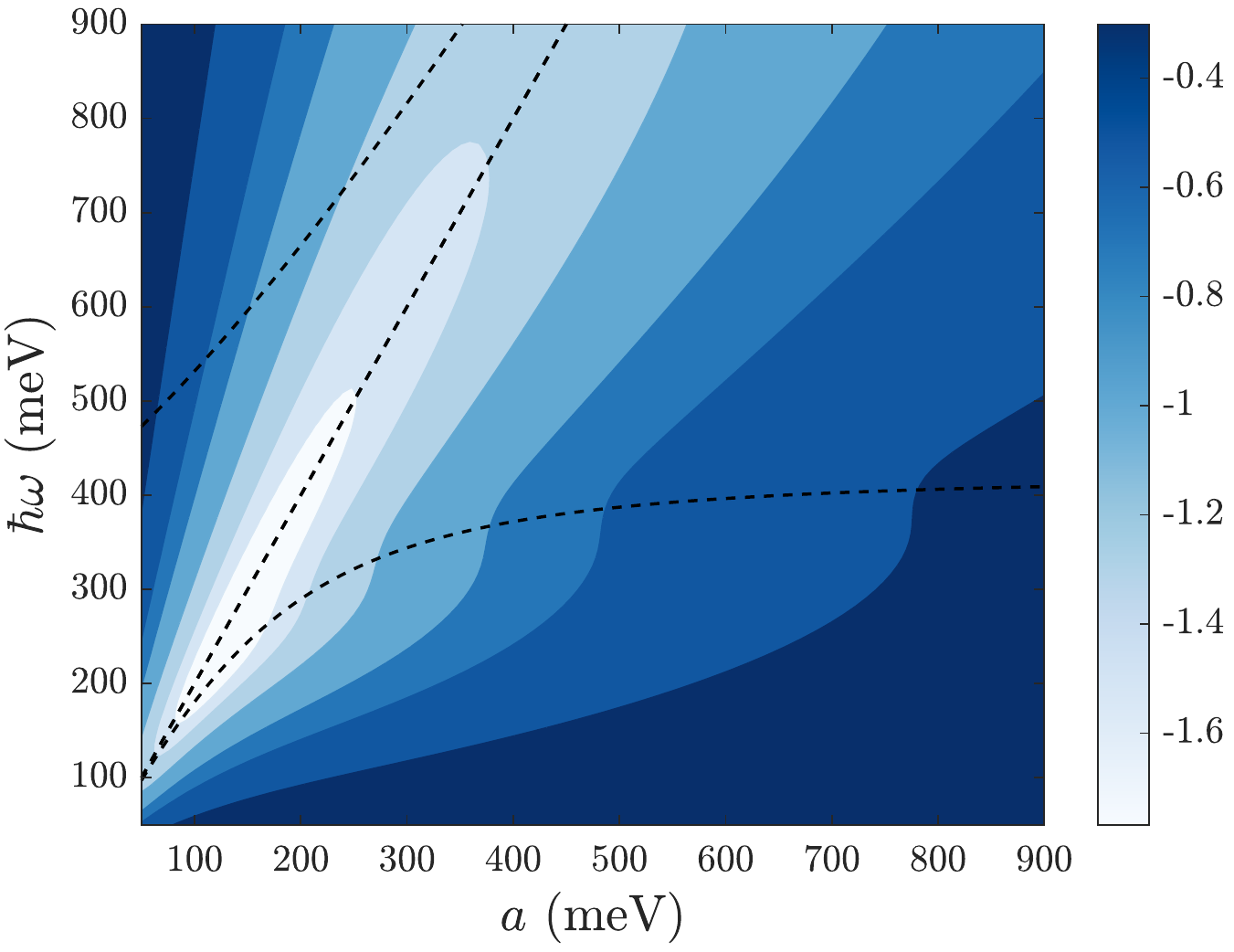}
    \caption{(Color online) \cite{colormap} Deviation of the second-order valley-polarization from perfect polarization $\text{log}_{10}(1-\mathcal{P}^{(2)})$ as a function of the external bias $a$ and central photon energy $\hbar\omega$ of the exciting Gaussian pulse. Lighter regions correspond to stronger valley-polarizations. From lightest to darkest, the contours correspond to $\mathcal{P}^{(2)}=0.97,0.95,0.9,0.8,0.7,$ and 0.5. The straight dashed line is the line $\hbar\omega=2a.$ The lower dashed curve is the band gap energy [Eq. (\ref{bandgap})], while the upper dashed curve is the energy of the next-lowest transition involving the higher-energy bands that we have neglected [Eq. (\ref{HEB})]. The pulse duration is $t_p=50$ fs, and the valley-polarization is evaluated long after the pulse has passed. Intraband scattering is neglected ($\gamma_{cc}(\mbfk)=0$), and the interband decoherence time is taken to be $\tau_0=30$ fs. The temperature is $300$ K and the chemical potential is $\mu=0.$}
    \label{VPNS}
\end{figure}

In this section, we examine the second-order valley-polarization when there is no intraband scattering, but where there is interband decoherence. Thus, we let $\gamma_{cc}(\mbfk)=0$ and $\gamma_{cv}(\mbfk)=1/\tau_0,$ where $\tau_0$ is a phenomenological decoherence time. We take $\tau_0=30$ fs and the pulse duration $t_p=50$ fs. We restrict ourselves to photon energies greater than 50 meV, because for lower energies, the pulse duration can be shorter than a single period. In Fig. \ref{VPNS} we plot $1-\mathcal{P}^{(2)},$ that is, the deviation of the valley-polarization from perfect polarization $(\mathcal{P}^{(2)}=1),$ on a logarithmic scale as a function of the external bias $a$ and the central photon energy $\hbar\omega$ of the exciting right-hand circularly polarized Gaussian pulse \footnote{The field amplitude $E_0$ does not need to be specified as it factors out in the evaluation of $\mathcal{P}^{(2)}.$ As long as the observation time $t_f$ is several times the pulse duration, $t_f$ also does not need to be specified since $n_V^{(2)}$ is time-independent for long $t_f$ when $\gamma_{cc}(\mbfk)=0$ [see Eq. (\ref{nsubV})].}.
Lighter colors correspond to stronger valley-polarizations. The strongest valley-polarizations are concentrated in the low-frequency---low-bias regime, along the line $\hbar\omega=2a$ (indicated by a straight dashed line). The valley-polarization falls off on either side of $\hbar\omega=2a.$ The valley-polarization degrades and broadens with increasing frequency and bias. The innermost contour corresponds to $\mathcal{P}^{(2)}>0.97.$ The two next-to-innermost contours correspond to $\mathcal{P}^{(2)}>0.95$ and 0.90 respectively. Over the parameter space considered in Fig. \ref{VPNS}, the valley-polarization ranges from $0.10\leq\mathcal{P}^{(2)}\leq 0.98.$ The optimal operating frequency-bias pair occurs for $(\hbar\omega,a)\approx(241,127)$ meV. Note however that both the optimal operating pair and the corresponding valley-polarization depend on the pulse duration and decoherence time.

Except at very low biases, the optimal operating frequencies do not coincide with the band gap energy (indicated by the lower dashed curve). In fact, the valley-polarization appears to be somewhat suppressed for frequencies resonant with the band gap energy. The upper dashed curve in Fig. \ref{VPNS} gives the energy of the next-lowest transition, involving the high-energy bands that we have neglected \footnote{In particular, this is the transition between either the high-energy valence band and the low-energy conduction band, or the low-energy valence band and the high-energy conduction band (see Ref. \cite{McCannTB}).}. Energies above this curve will induce significant transitions between bands other than the two low-energy bands we have considered. This curve is given explicitly by
\begin{equation}\label{HEB}
    \Delta E_\text{HB}(a) = (a^2+t_\perp^2)^{1/2}+a,
\end{equation}
which is strictly greater than $2a.$

We now examine in detail the origins of the main features of Fig. \ref{VPNS}, and in particular, why the valley-polarization is generally greatest along the line ${\hbar\omega=2a.}$ First, note that the first-order interband coherence $\rho_{cv}^{(1)}(\mbfk)$ is proportional to the carrier-field interaction $\boldsymbol{\xi}_{cv}(\mbfk)\cdot\mathbf{E}(t)$ [see Eq. (\ref{rhocv})]. If $\boldsymbol{\xi}_{cv}(\mbfk)\cdot\mathbf{E}(t)$ can be forced to zero in one valley but not the other, a strong valley-polarization is expected. For a right-hand circularly polarized field with central frequency $\omega,$ it can be easily shown using Eqs. (\ref{xiAB}) and (\ref{field}) that the carrier-field interaction is given by
\begin{equation}\label{dotproduct}
    \boldsymbol{\xi}_{cv}(\mbfk)\cdot\mathbf{E}(t)\approx iE(t)e^{-i(\omega t-\theta_k)}\big(A(a,k)\pm B(a,k)\big),
\end{equation}
where $A$ and $B$ are the real, positive functions that were introduced in Sec. \ref{Connection}, and where the approximation sign indicates that we are considering only the resonant contribution. Again, the plus and minus signs correspond to the $K$ and $K'$ valleys respectively. We see from Eq.~(\ref{dotproduct}) that in the $K$ valley, the interaction is proportional to the sum of the (positive) functions $A$ and $B,$ while in the $K'$ valley the interaction is proportional to the difference. When $A=B,$ Eq. (\ref{dotproduct}) amounts to a valley-contrasting optical selection rule in favor of the $K$ valley. Because $A$ and $B$ are $k$-dependent, the selection rule is in general not exact, except for very specific states within each valley (namely, states for which $A=B$). Note that if one uses light of the opposite helicity, the dominant piece of the carrier-field interaction is instead proportional to $(A\mp B),$ such that when $A=B$ the selection rule favors $K'.$

\begin{figure}
\includegraphics[width=\linewidth]{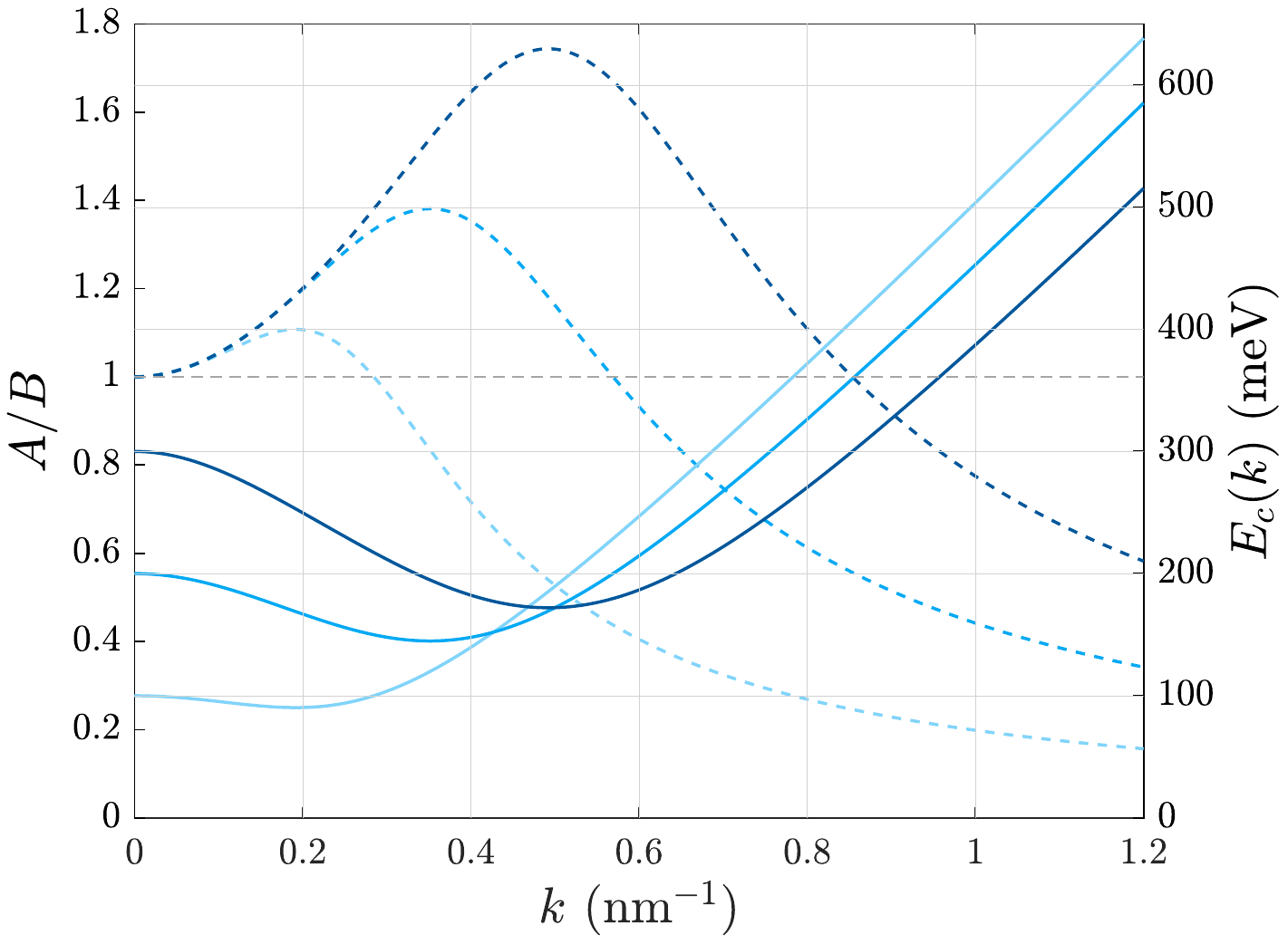}
\centering
\caption[AB Ratio]{$A/B$ ratio for several biases (dashed curves, leftmost axis), and corresponding conduction band energies $E_c(k)$ for the same biases (solid curves, rightmost axis) as a function of $k.$ From lightest to darkest, $a=100, 200,$ and 300 meV. The dashed horizontal line indicates $A/B=1.$}
\label{ABRatio}
\end{figure}

The ratio $A/B$ gives a measure of the ``exactness'' of the optical selection rule ($A/B=1$ when $A=B$). In Fig.~\ref{ABRatio}, we plot the quantity $A/B$ as a function of $k$ for three different biases (dashed curves), along with the corresponding conduction band energies, $E_c(k)$ (solid curves).
For all biases, $A/B=1$ at $k=0,$ indicating an exact selection rule. As one moves away from $k=0,$ the selection rules softens as the $A/B$ ratio deviates from unity. The $A/B$ ratio reaches its maximum at the band-minimum. The $A/B$ ratio then decays to zero as $k\rightarrow\infty.$ However, as the $A/B$ ratio decays, it passes through $A/B=1$ at some $k=k_1.$ In other words, for any given bias, the optical selection rule is exact at precisely two values of $k$: ${k=0}$ and ${k=k_1.}$ If we wish to achieve a strong valley-polarization, we should try to induce excitations at these very $k.$ However, as was discussed briefly in Sec. \ref{Connection}, both $A$ and $B$ vanish at $k=0.$ The carrier-field interaction [Eq. (\ref{dotproduct})] therefore vanishes at $k=0$ (for both $K$ and $K'$), and so there is no carrier injection at $k=0.$ Therefore, in what follows, we focus our discussion on the states near $k=k_1.$

In Fig. \ref{ABRatio}, we saw that the $A/B$ ratio peaks at the band-minimum. In fact, one can show that
\begin{equation}\label{ABaE}
    \frac{A}{B}=\frac{a}{E_c(k)}.
\end{equation}
Thus, $A/B=1$ when $E_c(k)=a,$ which occurs precisely at $k=0$ and $k=k_1,$ where
\begin{equation}\label{k1}
    k_1=\frac{4a}{3a_0t_{\|}}=\frac{2a}{\hbar v_f}.
\end{equation}
To target these states, we must tune the frequency of the exciting field such that it is resonant with interband transitions at $k=k_1.$ Due to electron-hole symmetry, the transition energy at $k_1$ is simply $2E_c(k_1),$ and so the optimal operating frequency for a particular bias is expected to be given by
\begin{equation}\label{hw2a}
    \hbar\omega=2E_c(k_1)=2a,
\end{equation}
which explains why the optimal operating frequency-bias pairs in Fig. \ref{VPNS} lie along the line $\hbar\omega=2a.$ In monolayer graphene with a staggered sublattice potential, one finds an exact selection rule at the Dirac points ($k=0$) \cite{yao2008valley}. In contrast, in biased bilayer graphene, because $A$ and $B$ both vanish at exactly $\mathbf{K}$ and $\mathbf{K'},$ there are no carriers injected at $k=0$ and therefore there is no valley-contrasting optical selection rule at $k=0.$ Rather, the optical selection rule is found along a ring of states with radius $k=k_1$ surrounding the Dirac points. The existence of such a $k$ value seems to have gone unnoticed in the literature. However, evidence for this result can be seen in Fig. 2 of Yao \textit{et al.} \cite{yao2008valley}.

The valley-polarization is robust to deviations from $\hbar\omega=2a.$ In Fig. \ref{VPNS}, the innermost contour corresponded to $\mathcal{P}^{(2)}>0.97.$ Consider operating near the optimal frequency and bias $(\hbar\omega,a)=(241,127)$ meV, for which $\mathcal{P}^{(2)}=0.98.$ For fixed $\hbar\omega=241$ meV, external biases within the range $110<a<161$ meV yield $\mathcal{P}^{(2)}>0.97.$ For fixed $a=127$ meV, central photon energies within $198<\hbar\omega<284$ meV yield $\mathcal{P}^{(2)}>0.97.$ Similarly, the third-to-innermost contour corresponded to $\mathcal{P}^{(2)}>0.90.$ For fixed $\hbar\omega=241$ meV, $84<a<230$ meV yields $\mathcal{P}^{(2)}>0.90$. For fixed $a=127$ meV, $146<\hbar\omega<396$~meV yields $\mathcal{P}^{(2)}>0.90$. Thus, deviations of several 10s of meV in either frequency or bias from the optimal operating pair still yield valley-polarizations well-over 90\%. If one cannot target the optimum precisely, it is better to err towards larger frequencies and biases.

In Fig. \ref{VPNS}, the optimal operating frequencies coincide with the band gap for small biases. This is because the band gap energy $\Delta E(a)\rightarrow 2a$ in the limit of small $a$ [see Eq. (\ref{bandgap})]. For more moderate biases, we see from Fig. $\ref{VPNS}$ that the valley-polarization appears somewhat reduced at frequencies close to the band gap energy. This is because for any given bias, the $A/B$ ratio reaches its maximum at the band edge and therefore in general leads to a poor valley-polarization [see Fig. \ref{ABRatio} or Eq. (\ref{ABaE})]. For frequencies less than the band gap energy, carriers are still predominantly excited at the band edge due to the finite bandwidth of the Gaussian pulse. Only once the central frequency exceeds the band gap energy do carriers away from the band edge begin to dominate the response. The valley-polarization is therefore not reduced near the band gap, rather, the valley-polarization ``stalls'' at the band gap energy as one sweeps upwards in frequency. This result is in agreement with Fig. 3 of Ref. \cite{abergel}, where a strong valley-polarization was calculated for photon energies close to resonance with the band gap energy $\Delta E(a)\approx 2a$ for a small bias of $2a=20$ meV.

In Fig. \ref{VPNS}, the stripe of optimal operating frequency-bias pairs is seen to broaden with increasing $a$ and $\hbar\omega$. This can be understood as follows. The energy-derivative of the $A/B$ ratio at $k=k_1$ is a measure of the sensitivity of the optical selection rule to small deviations in photon energy from $\hbar\omega=2a.$ One may easily verify that
\begin{equation}
    \frac{d}{dE_c(k)}\left(\frac{A}{B}\right)\bigg\rvert_{k=k_1}=-\frac{1}{a}.
\end{equation}
Thus, as the bias is increased, the system becomes less sensitive to small deviations from the optimal operating configuration and the valley-polarization broadens.

For a couple of reasons, one can never achieve a perfect valley-polarization. First, targeting states for which $A/B=1$ exclusively is impossible. Even when the pulse duration is very long, the decoherence time results in linewidth broadening. This means that even when the central photon energy of the pulse is $\hbar\omega=2a,$ carriers will be excited not just at $k=k_1,$ but also in the \textit{vicinity} of $k=k_1$ where $A\neq B$ and the valley-contrasting optical selection rule is inexact. In Fig. \ref{VPNS}, we observed that the valley-polarization degrades along $\hbar\omega=2a$ with increasing frequency and bias. This degradation is only observed in the presence of finite decoherence times. Second, the valley-polarization is limited by the presence of the anti-resonant piece of the carrier field interaction. In contrast to the resonant term [Eq. (\ref{dotproduct})], the anti-resonant term is proportional to $(A\mp B),$ meaning that the selection rule can never be truly exact. That is, when $A=B$ the resonant term vanishes but the anti-resonant term does not. For the particular pulse duration and decoherence time chosen in Fig. \ref{VPNS}, the difference between the valley-polarization obtained if one includes both the resonant and anti-resonant terms, and the valley-polarization obtained when one includes just the resonant terms, can be as large as $\Delta\mathcal{P}^{(2)}=0.05$ along the line $\hbar\omega=2a$ (over the parameter space considered). The greatest differences occur at the edges of the parameter space. Close to the optimal operating frequency-bias pair, the difference is significantly smaller, about $\Delta\mathcal{P}^{(2)}=0.01.$ Since the valley-polarization can reach up to $\mathcal{P}^{(2)}=0.98,$ a difference of $0.01$ is significant. The anti-resonant term must therefore be included to obtain accurate results. Taken together, a finite decoherence time and the presence of the anti-resonant term ensures that the valley-polarization is never complete. However, as we shall discuss in Sec. \ref{Effects}, the impact of these two effects can be reduced by increasing the decoherence time and pulse duration. 

In the ideal limit of infinite decoherence times and infinite pulse durations, with $t_p\gg \tau_0,$ one can show that the valley-polarization approaches unity when operating at $\hbar\omega=2a.$ To see this, consider Eq. (\ref{nsubV}). In the ideal limit, the exponentials $\text{exp}(\beta_\mp)$ approach delta functions centered on $\omega_{cv}(\mbfk)=\mp \omega.$ Since $\omega>0$ and $\omega_{cv}(\mbfk)=2E_c(k)/\hbar>0,$ the anti-resonant ($+$) term contributes nothing to the carrier density, leaving only the resonant ($-$) term. If $\hbar\omega=2a,$ then $A=B$ and the resonant term vanishes in $K',$ resulting in perfect valley-polarization.

\subsection{With intraband scattering}\label{IntrabandScattering}
We now improve upon our model by accounting for intraband scattering processes which act to reduce the valley-polarization. Intervalley scattering (IVS) is a process in which an electron in (say) the $K$ valley scatters to $K',$ degrading the valley-polarization. As was discussed in the beginning of Sec. \ref{Results}, the relevant relaxation pathways for photoexcited carriers in bilayer graphene are carrier-carrier and carrier-phonon scattering. Due to momentum conservation, IVS via carrier-carrier interactions is very weak \cite{Andreas}. We find it crucial, however, to account for carrier-phonon interactions as IVS via optical phonons can significantly reduce the valley-polarization when the frequency of the exciting pulse is large.

Carrier-phonon scattering in monolayer graphene (MLG) has been studied extensively \cite{Ballistic,OpticalDom,Lazzeri}. Unfortunately, the literature on carrier-phonon scattering in bilayer graphene (BLG), and in particular in biased bilayer graphene (BBLG), is somewhat lacking and there is no clear consensus on the dominant scattering modes. In this work, we treat scattering to be the same as in MLG where optical phonons dominate at room temperature \cite{OpticalDom}. This is certainly an approximation, for it is not clear whether scattering in BBLG is similar to scattering in MLG. For instance, a 2011 study has suggested that, in contrast to MLG, low-energy acoustic phonons dominate in \textit{unbiased} BLG, while optical phonon scattering is highly suppressed \cite{EPCBLG}. However, it is unclear if this result holds for a biased bilayer: a 2015 study found that optical phonons \textit{are} the dominant scattering mode in BBLG at biases above $a=150$ meV \cite{suomi}. If it turns out that acoustic phonons play an important role in BBLG transport, our calculations can be easily modified to account for them.

\begin{figure}
    \centering
    \includegraphics[width=\linewidth]{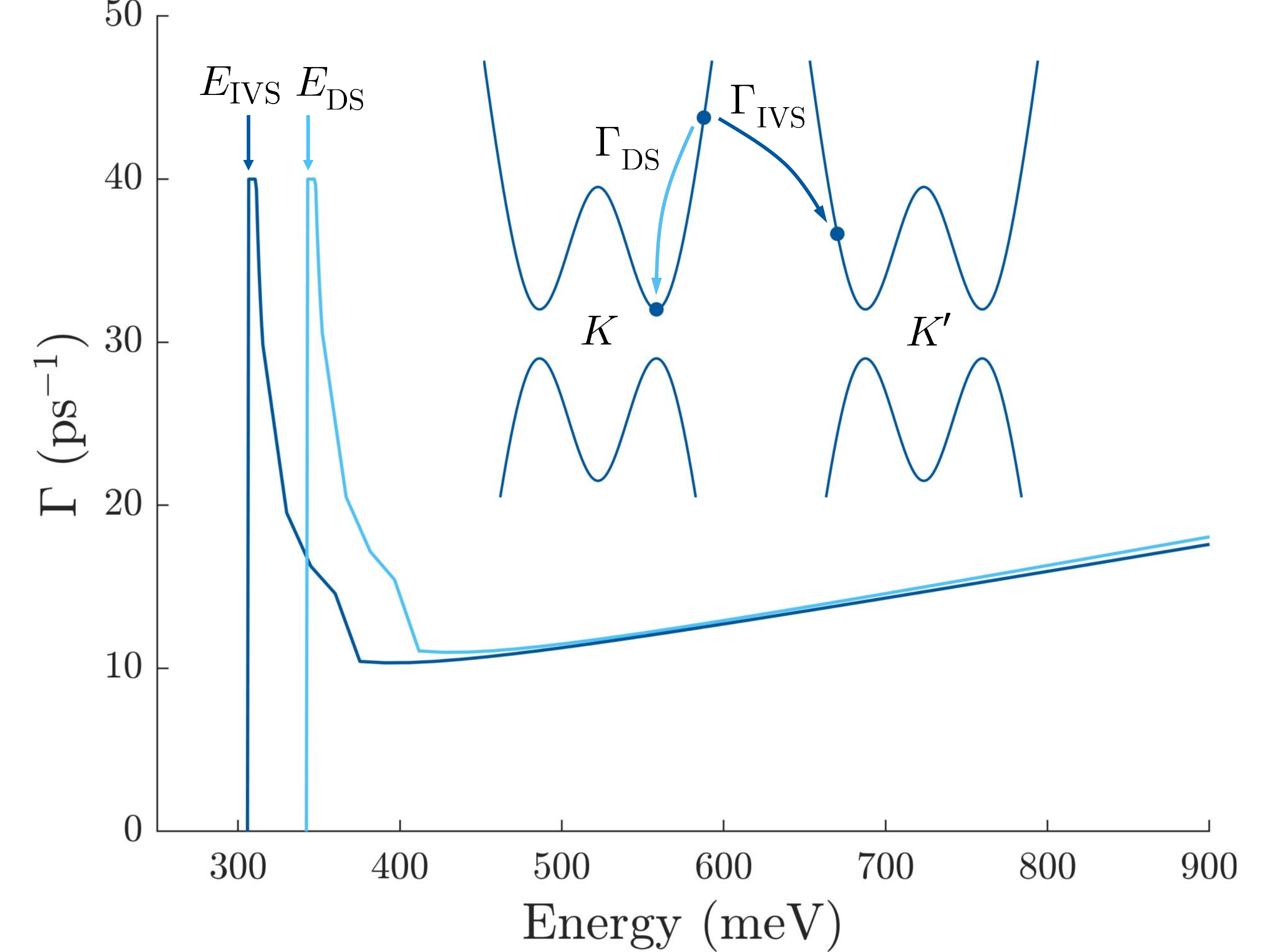}
    \caption[Scattering Schematic]{Scattering rates $\Gamma_\text{IVS}$ (dark) and $\Gamma_\text{DS}$ (light) as functions of energy for an example bias of $a=205$ meV. The sharp onsets of $\Gamma_\text{IVS}$ and $\Gamma_\text{DS}$ occur at energies $E_\text{IVS}(a)$ and $E_\text{DS}(a),$ respectively. The scattering rates are cut off at a maximum value of $\Gamma=\Gamma_\text{max}=40$ ps$^{-1}$ (see text). The inset schematically depicts the corresponding scattering processes.}
    \label{ScatteringSchematic}
\end{figure}

To incorporate intraband scattering into our model, we adopt the formalism of Ref. \cite{luke}. In this approach, scattering is treated microscopically, rather than phenomenologically, as we have done up to this point. Beginning with a Fr\"ohlich Hamiltonian and making the second-order Born-Markov approximation, one obtains scattering rates $\Gamma(\mbfk)$ that appear analogously to $\gamma_{cc}(\mbfk)$ in Eq. (\ref{rhonm}). Electron-phonon scattering can therefore be incorporated in our model by simply replacing $\gamma_{cc}(\mbfk)$ with $\Gamma(\mbfk)$ in Eq. (\ref{nsubV}). At room temperature, the thermal population of optical phonons is very small, and so the dominant inelastic scattering process in MLG is optical phonon emission. Optical phonons with crystal momenta near the $K$ points are the only phonons capable of assisting in intervalley scattering. However, optical phonons near the $\Gamma$ point can assist in \textit{intra}valley ``down-scattering'' (DS), a process which leaves the scattered carrier in its original valley. IVS and DS processes are depicted schematically in the inset of Fig. \ref{ScatteringSchematic}.
It is important to include DS along with IVS because DS will limit the number of carriers available to IVS. The dominant contribution to IVS comes from the TO mode at the $K$ points, while the dominant contribution to DS comes from the degenerate TO and LO modes at $\Gamma$ \cite{Piscanec}. All other phonon modes are neglected, and all other elastic/quasi-elastic scattering processes are accounted for through the phenomenological decoherence time $\tau_0.$ 

The scattering-out rates for conduction band electrons due to $K$- and $\Gamma$-phonon \mbox{emission are} \cite{luke}
\begin{align*}
\Gamma_\text{IVS}(\mathbf{k})&=\frac{2 \pi}{\hbar N} \sum_{\mathbf{q}}\frac{1}{2} g_{K}^{2}(1-\rho_{c c}^{(0)}(\mathbf{k}-\mathbf{q}))\left(n_{K}+1\right)\\&\qquad\qquad\times \delta\left[E_{c}(\mathbf{k}-\mathbf{q})-E_{c}(\mathbf{k})+\hbar \omega_{K}\right],\numberthis{}\label{GammaIVS}\\
\Gamma_\text{DS}(\mathbf{k})&=\frac{2 \pi}{\hbar N} \sum_{\mathbf{q}} 2 g_{\Gamma}^{2}(1-\rho_{c c}^{(0)}(\mathbf{k}-\mathbf{q}))\left(n_{\Gamma}+1\right) \\&\qquad\qquad\times\delta\left[E_{c}(\mathbf{k}-\mathbf{q})-E_{c}(\mathbf{k})+\hbar \omega_{\Gamma}\right], \numberthis{}\label{GammaDS}
\end{align*}
where $N$ is the number of unit cells, and where we have approximated the angle for IVS events to be $60\degree$ (the angle between the $\mathbf{K}$ and $\mathbf{K}'$ points, as measured from the zone-center). The quantities $g_K^2$ and $g_\Gamma^2$ are constants, but there is some debate in the literature as to their precise values \cite{Piscanec,BorysMLG,RamanNoodles,debated}. Following Ref. \cite{Lazzeri}, we take $g_K^2=0.2098$ eV$^2$ and $g_\Gamma^2=0.0558$ eV$^2,$ noting that their precise values will have little effect on our conclusions. Since the high-energy phonon dispersion of BLG (and BBLG) is very similar to MLG \cite{BLGPhononDisp,MatTodayProc}, we follow Ref. \cite{luke} and take the phonons to be dispersionless near the $K$ and $\Gamma$ points with constant frequencies $\hbar\omega_K=160$~meV and $\hbar\omega_\Gamma=196$ meV. We treat the phonons as thermal baths at $300$ K so that the phonon populations $n_{K}$ and $n_{\Gamma}$ are given by Bose-Einstein distributions at energy $\hbar\omega_{K}$ and $\hbar\omega_\Gamma$ respectively. Since $E_c(\mbfk)=E_c(k)$ and $\rho_{cc}^{(0)}(\mbfk)=f_c(k),$ the scattering-out rates are isotropic and we may write $\Gamma_\text{IVS}(\mbfk)=\Gamma_\text{IVS}(k)$ and $\Gamma_\text{DS}(\mbfk)=\Gamma_\text{DS}(k).$ We now let
\begin{equation}
    \gamma_{cc}(\mbfk)=\Gamma_\text{IVS}(k)+\Gamma_\text{DS}(k).
\end{equation} 
We also modify the interband decoherence rate according to $\gamma_{cv}(\mbfk)=1/\tau(k),$ where
\begin{equation}\label{tauk}
    \frac{1}{\tau(k)}=\frac{1}{\tau_0}+\frac{\gamma_{cc}(\mbfk)}{2},
\end{equation}
and where the factor of $1/2$ arises from the usual relationship between decoherence and population decay rates.

In Fig. \ref{ScatteringSchematic}, we plot $\Gamma_\text{IVS}(k)$ and $\Gamma_\text{DS}(k)$ as functions of \textit{energy} for an example bias of $a=205$ meV. As is demonstrated in the figure, the Dirac delta functions in Eqs. (\ref{GammaIVS}) and (\ref{GammaDS}) ensure that intraband optical-phonon-scattering is forbidden unless an excited electron can afford to lose the energy of a phonon and remain in the conduction band. In other words, $\Gamma_\text{IVS}(k)$ and $\Gamma_\text{DS}(k)$ are, respectively, strictly zero unless 
\begin{align}
    E_c(k)\geq\hbar\omega_{K/\Gamma}+\Delta E(a)/2 \equiv E_\text{IVS/DS}(a), \label{onset}
\end{align}
where $\Delta E(a)$ is the band gap. Since ${\hbar\omega_K<\hbar\omega_\Gamma,}$ IVS becomes energetically possible before DS. Due to electron-hole symmetry, the photon energy required to excite an electron from the valence band to an energy $E_\text{IVS}(a)$ or $E_\text{DS}(a)$ in the conduction band is given by $\hbar\omega=2E_\text{IVS}(a)$ or $\hbar\omega=2E_\text{DS}(a),$ respectively. A second consequence of the delta functions is that $\Gamma_\text{IVS}(k)$ and $\Gamma_\text{DS}(k)$ diverge when $E_c(k)=E_\text{IVS}(a)$ and $E_c(k)=E_\text{DS}(a),$ respectively. Away from these energies, the scattering rates are well-behaved and on the order of a few 10s of ps$^{-1},$ which is consistent with rates found in Refs. \cite{OpticalDom,ShotNoise}. Due to the approximations we made when evaluating the carrier density (Sec. \ref{Calculating}), a divergent intraband scattering rate results in numerical difficulties unless the observation time $t_f$ is very large. More precisely, large values of $\gamma_{cc}(\mbfk)$ push the peak of the integrand of Eq. (\ref{precompl}) towards $t=\infty.$ In requiring that the integrand decay sufficiently to zero by $t=t_f,$ a divergent $\gamma_{cc}(\mbfk)$ forces us longer and longer $t_f$ at which to evaluate the carrier density. To address this problem, we limit $\Gamma_\text{IVS}(k)$ and $\Gamma_\text{DS}(k)$ to a maximum value $\Gamma_\text{max}=40$ ps$^{-1}$ (see Fig. \ref{ScatteringSchematic}). With this value of $\Gamma_\text{max},$ for pulse durations $t_p$ on the order of 10s to a few 100s of fs, and for typical decoherence times, we obtain convergence for $t_f=5t_p.$ We find that although changing the value of $\Gamma_\text{max}$ affects the high-frequency results (where scattering is strong), it has very little effect in the optimal operating region.

\subsubsection{Intervalley scattering}
To begin with, let us simplify things and neglect down-scattering by setting $\Gamma_\text{DS}(k)=0,$ such that $\gamma_{cc}(\mbfk)=\Gamma_\text{IVS}(k).$ Because we treat the $K$ and $K'$ valleys as disconnected in $k$-space, and we do not account for carriers scattering-in, we cannot keep track of IVS explicitly. That is, $\gamma_{cc}(\mbfk)$ simply acts as a population decay rate for electrons in each valley \textit{independently} [see Eq. (\ref{nsubV})]. We therefore require a systematic way to keep track of the number of carriers which IVS from each valley. Once we have that, we can then add those carriers into the opposing valley before computing the valley-polarization.

To this end, we calculate the carrier density in each valley twice: once subject to decay from IVS, and a second time without population decay. If IVS is the only scattering process, then the difference between these two quantities gives an estimate of the density of carriers that have intervalley scattered. Concretely, let $n_K^\text{IVS}$ be the carrier density injected into the $K$ valley at time $t=t_f$, calculated using Eq. (\ref{nsubV}) with $\gamma_{cc}(\mbfk)=\Gamma_\text{IVS}(k).$ As before, $n^{(2)}_K$ is the carrier density injected into $K,$ calculated using Eq. (\ref{nsubV}) with $\gamma_{cc}(\mbfk)=0.$ A summary of these quantities are given in Table \ref{scatteringterms}. The quantity
\begin{equation}\label{deltaK}
    \Delta n_{K'}^\text{IVS}=n^{(2)}_{K}-n_{K}^\text{IVS}
\end{equation}
gives the density of carriers intervalley scattered from $K$ to $K'$ (the subscript on $\Delta n_{K'}^\text{IVS}$ denotes the final valley). The carrier density in the $K'$ valley is then modified according to \begin{equation}\label{nKprime}
    \widetilde{n}^{(2)}_{K'}=n_{K'}^\text{IVS}+\Delta n_{K'}^\text{IVS}.
\end{equation}
In other words, the carrier density in $K'$ is given by the carrier density \textit{remaining} in $K'$ (after allowing for population decay from IVS), plus the carriers that have intervalley scattered from $K$ to $K'.$ The carrier density in the $K$ valley is obtained by interchanging $K \leftrightarrow K'$ in Eqs.~(\ref{deltaK}) and (\ref{nKprime}).

\renewcommand{\arraystretch}{1.5}
\begin{table}
 \caption{Summary of the notation used for the carrier densities in the $K$ valley and their associated scattering rates. Each density is calculated using Eq. (\ref{nsubV}) with $\gamma_{cc}(\mbfk)$ as indicated and $\gamma_{cv}(\mbfk)$ according to Eq. (\ref{tauk}). The $K'$ densities are defined analogously.}
    \label{scatteringterms}
\begin{center}
    \begin{tabular}{|c|c|c|c|c|}
        \hline
          &  $n_K^{(2)}$ & $n_K^{\text{IVS}}$ & $n_K^{\text{DS}}$ & $n_K^{\text{IVS+DS}}$\\
      \hline
         $\gamma_{cc}(\mbfk)$ & 0 & $\Gamma_\text{IVS}$ & $\Gamma_\text{DS}$ & $\Gamma_\text{IVS}$+$\Gamma_\text{DS}$\\
      \hline
    \end{tabular}
\end{center}
\end{table}

\begin{figure*}
    \centering
    \subfloat[\label{scattering:IVS}]{
    \includegraphics[width=0.49\linewidth]{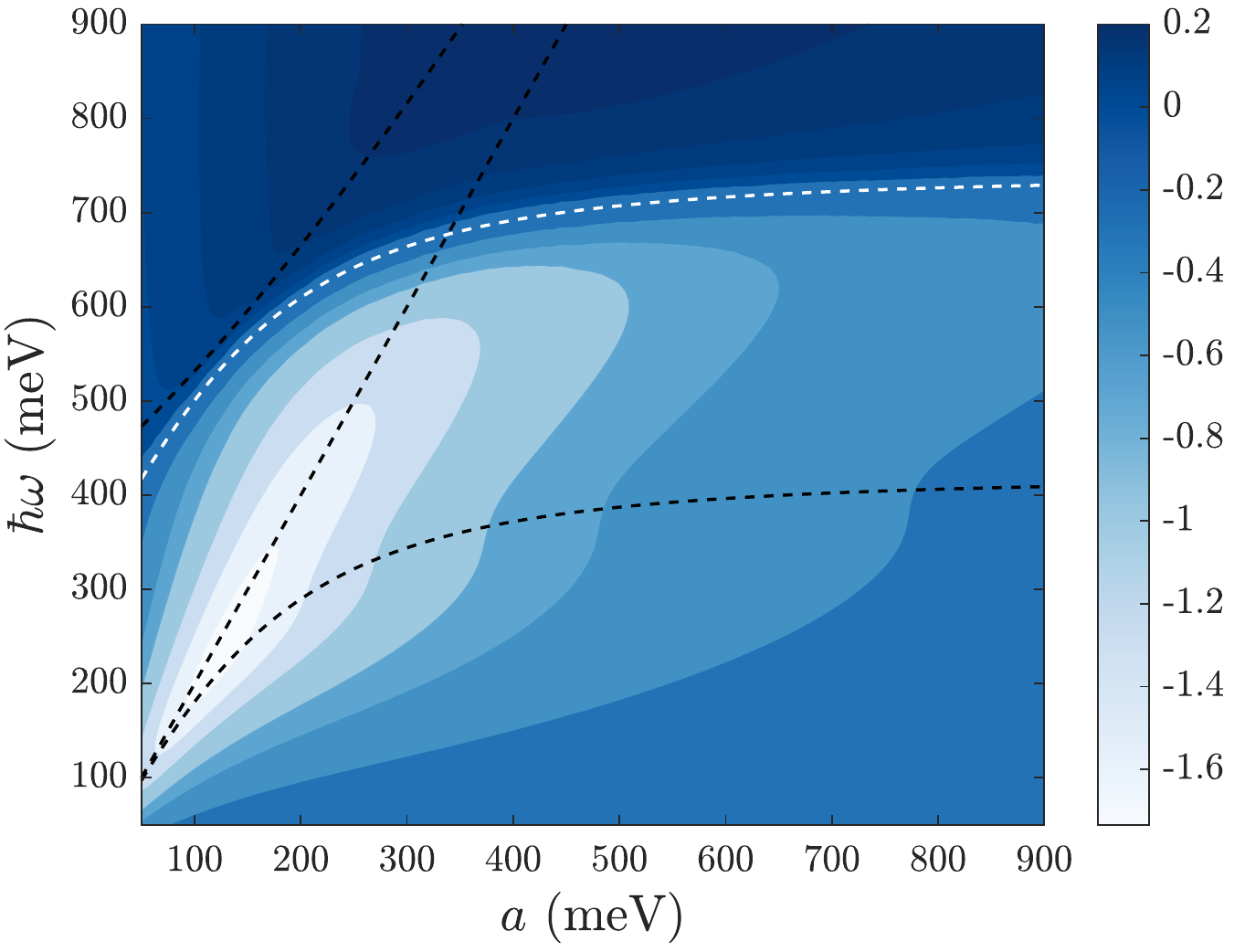}%
        }
    \subfloat[\label{scattering:DS}]{%
    \includegraphics[width=0.49\linewidth]{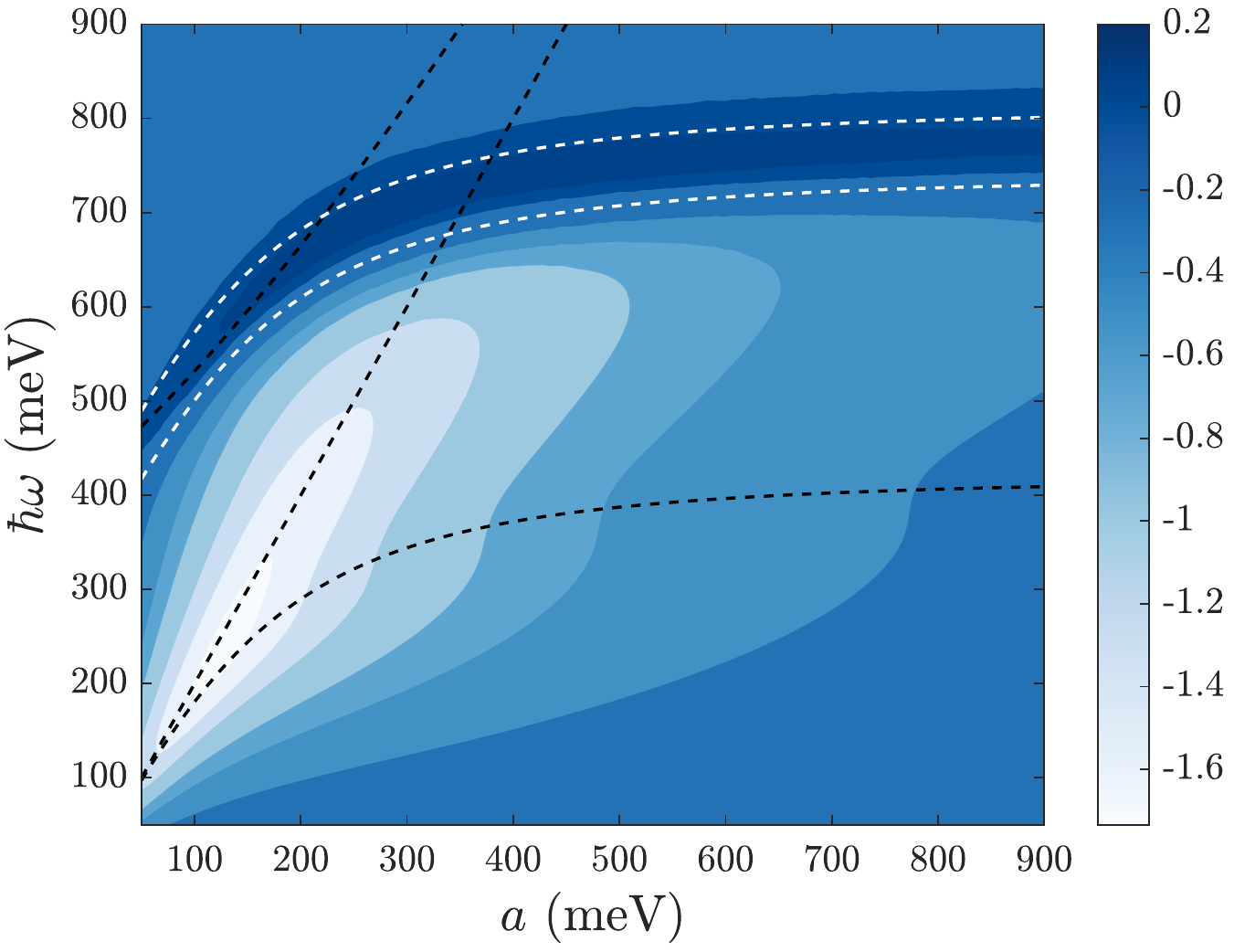}%
        }
        \caption{(Color online) Deviation of the second-order valley-polarization from perfect polarization $\text{log}_{10}(1-\mathcal{P}^{(2)})$ as a function of the external bias $a$ and central photon energy $\hbar\omega$ of the exciting Gaussian pulse with the inclusion of (a) intervalley scattering and (b) both intervalley scattering and down-scattering. The dashed black curves are as in Fig. \ref{VPNS}. In (a), the dashed white curve is the onset of intervalley scattering, given by $\hbar\omega=2E_\text{IVS}(a)$ [Eq. (\ref{onset})]. In (b), the lower dashed white curve is the same curve as in (a), while the upper dashed white curve is the onset of down-scattering, given by $\hbar\omega=2E_\text{DS}(a)$. From lightest to darkest, the contours in (a) and (b) correspond to $\mathcal{P}^{(2)}=0.975,0.95,0.9,0.8,0.7,0.6,0.5,0,-0.15,-0.3,-0.45$ and $-0.6.$ Note that the final three contours are not visible in (b) because the data does not extend to these values. In both (a) and (b), the phenomenological decoherence time is $\tau_0=30$ fs, the pulse duration is $t_p=50$ fs, and the valley-polarization is evaluated at $t_f=250$ fs. The temperature is $300$ K and the chemical potential is $\mu=0.$}\label{scattering}
\end{figure*}

The valley-polarization is modified to
\begin{equation}\label{Pmod}
    \mathcal{P}^{(2)}=\frac{\widetilde{n}^{(2)}_K-\widetilde{n}^{(2)}_{K'}}{\widetilde{n}^{(2)}_K+\widetilde{n}^{(2)}_{K'}},
\end{equation}
which we plot as a function of the external bias and central photon energy in Fig.~\subref*{scattering:IVS} under the same conditions as Fig. \ref{VPNS} ($t_p=50$ fs, $\tau_0=30$ fs, $T=300$ K, $\mu=0$).
As was the case in Fig. \ref{VPNS}, the strongest valley-polarizations are concentrated in the low-frequency---low-bias regime, along $\hbar\omega=2a.$ The most obvious difference between Fig.~\subref*{scattering:IVS} and Fig.~\ref{VPNS} is the emergence of a dark blue region of poor valley-polarization in the high-frequency portion of the parameter space. This is due to the presence of intervalley scattering, the onset of which is indicated by the dashed white curve corresponding to $\hbar\omega=2E_\text{IVS}(a).$ Notice that the onset of IVS intersects the line $\hbar\omega=2a,$ meaning that IVS effectively cuts off high frequencies from the parameter space of optimal operating frequency-bias pairs. Unfortunately, high frequencies may be desirable due to hot-carrier multiplication, which would result in a larger population imbalance and thus a stronger valley-polarization \cite{hotcarriers,hotcarrierAndreas}. If low-energy acoustic phonon scattering turns out to play an important role in BBLG, one would expect to see an onset curve analogous to the one in Fig.~\subref*{scattering:IVS}, but shifted towards lower photon energies. In comparison to Fig. \ref{VPNS}, the optimal operating frequency-bias pair is shifted slightly towards lower frequencies, $(\hbar\omega,a)=(236,126)$~meV, but a very strong valley-polarization of $\mathcal{P}^{(2)}=0.98$ is retained. The difference is slight because the optimal operating region is largely unaffected by the presence of intervalley scattering, but nonzero because the pulse excites over a broad energy range. In Fig.~\subref*{scattering:IVS}, the valley-polarization is evaluated long after the exciting pulse has passed (${t_f=5t_p=250}$~fs).

At photon energies greater than the onset of IVS, the valley-polarization can become negative, meaning that the system is valley-polarized in favor of $K'$ electrons. In fact, the five darkest shaded regions in Fig.~\subref*{scattering:IVS} correspond to $\mathcal{P}^{(2)}\leq 0.$ This is because the IVS rate is so strong in this region that more carriers intervalley scatter from $K$ to $K'$ than remain in the $K$ valley. This result suggests that it may be possible to achieve an inverse valley-polarization, that is, a valley-polarization in favor $K'$ electrons, even though the exciting pulse is right-hand circularly polarized. We emphasize that a significant fraction of this region of inverse valley-polarization lies above the upper black dashed curve $\Delta E_\text{HB}(a)$ and therefore should be taken with a grain of salt [Eq. (\ref{HEB})].

By neglecting down-scattering thus far, the results presented in Fig.~\subref*{scattering:IVS} give an underestimate of the valley-polarization. Carriers that DS contribute to the valley-polarization in the same way as if they had not scattered at all because they remain in their original valley. However, down-scattered carriers will likely have scattered into states below the energy threshold $E_\text{IVS}(a)$ required to IVS. In other words, DS acts to limit IVS by reducing the number of carriers available to IVS. We are again faced with the problem of keeping track of these carriers given that $\gamma_{cc}(\mbfk)$ simply acts as a population decay rate.

\subsubsection{Down-scattering}
To achieve a better estimate of $\mathcal{P}^{(2)},$ we must adjust Eqs. (\ref{deltaK}) and (\ref{nKprime}) to account for down-scattering. This essentially amounts to keeping track of another contribution to the carrier density in each valley. In the exact same way we estimated the density of carriers that intervalley scattered, let us define
\begin{equation}\label{deltaK2}
    \Delta n_{K'}^\text{DS}=n^{(2)}_{K'}-n_{K'}^\text{DS}
\end{equation}
to be the density of carriers scattered \textit{within} the $K'$ valley (again, the subscript on $\Delta n_{K'}^\text{DS}$ denotes the final valley). Analogously, $n_{K'}^{\text{DS}}$ is the carrier density injected into $K',$ calculated using Eq. (\ref{nsubV}) with $\gamma_{cc}(\mbfk)=\Gamma_\text{DS}(k)$ (see Table. \ref{scatteringterms}). The carrier density in the $K'$ valley becomes \begin{equation}\label{nKprime2}
    \widetilde{n}^{(2)}_{K'}=n_{K'}^\text{IVS+DS}+\Delta n_{K'}^\text{IVS}+\Delta n_{K'}^\text{DS}.
\end{equation}
Here, $n_{K'}^\text{IVS+DS}$ is the carrier density obtained by allowing for decay from both IVS and DS, that is, for $\gamma_{cc}(\mbfk)=\Gamma_\text{IVS}(k)+\Gamma_\text{DS}(k).$ In other words, the carrier density in the $K'$ valley is given by the carrier density remaining in $K'$ (after allowing for population decay from both IVS and DS), plus the carriers that IVS from $K$ to $K',$ plus the carriers that DS within $K'.$

In Fig.~\subref*{scattering:DS}, we plot Eq. (\ref{Pmod}) for $\mathcal{P}^{(2)},$ modified to include down-scattering according to the redefinitions above, and under the same conditions as Fig.~\subref*{scattering:IVS}. The general features that were observed in Fig~\subref*{scattering:IVS} are unchanged. In comparison to Fig.~\subref*{scattering:IVS}, the valley-polarization recovers somewhat after the onset of down-scattering, which is indicated by the upper dashed white curve corresponding to $\hbar\omega=2E_\text{DS}(a)$. However, the valley-polarization does not recover significantly enough to make this regime attractive for valleytronics. We therefore conclude that, even when accounting for DS, one should operate in the low-frequency---low-bias regime before the onset of IVS. The optimal operating frequency-bias pair is unchanged from the previous calculation: $(\hbar\omega,a)=(236,126)$ meV, $\mathcal{P}^{(2)}=0.98.$ Due to our approximate scheme for estimating carrier scattering, Fig.~\subref*{scattering:DS} simultaneously overestimates the number of carrier that IVS and the number of carriers that DS. Overestimating the number of carriers that IVS (DS) tends to reduce (increase) the valley-polarization. Because these two effects act in opposition, it difficult to discern whether Fig.~\subref*{scattering:DS} provides an overestimate or an underestimate of the valley-polarization. However, we believe it provides a better estimate than Fig.~\subref*{scattering:IVS}, which should be considered a worst-case scenario.

\subsection{Pulse duration and decoherence time}\label{Effects}
\begin{figure*}
    \centering
    \subfloat[\label{effects:varytp}]{
    \includegraphics[width=0.5\linewidth]{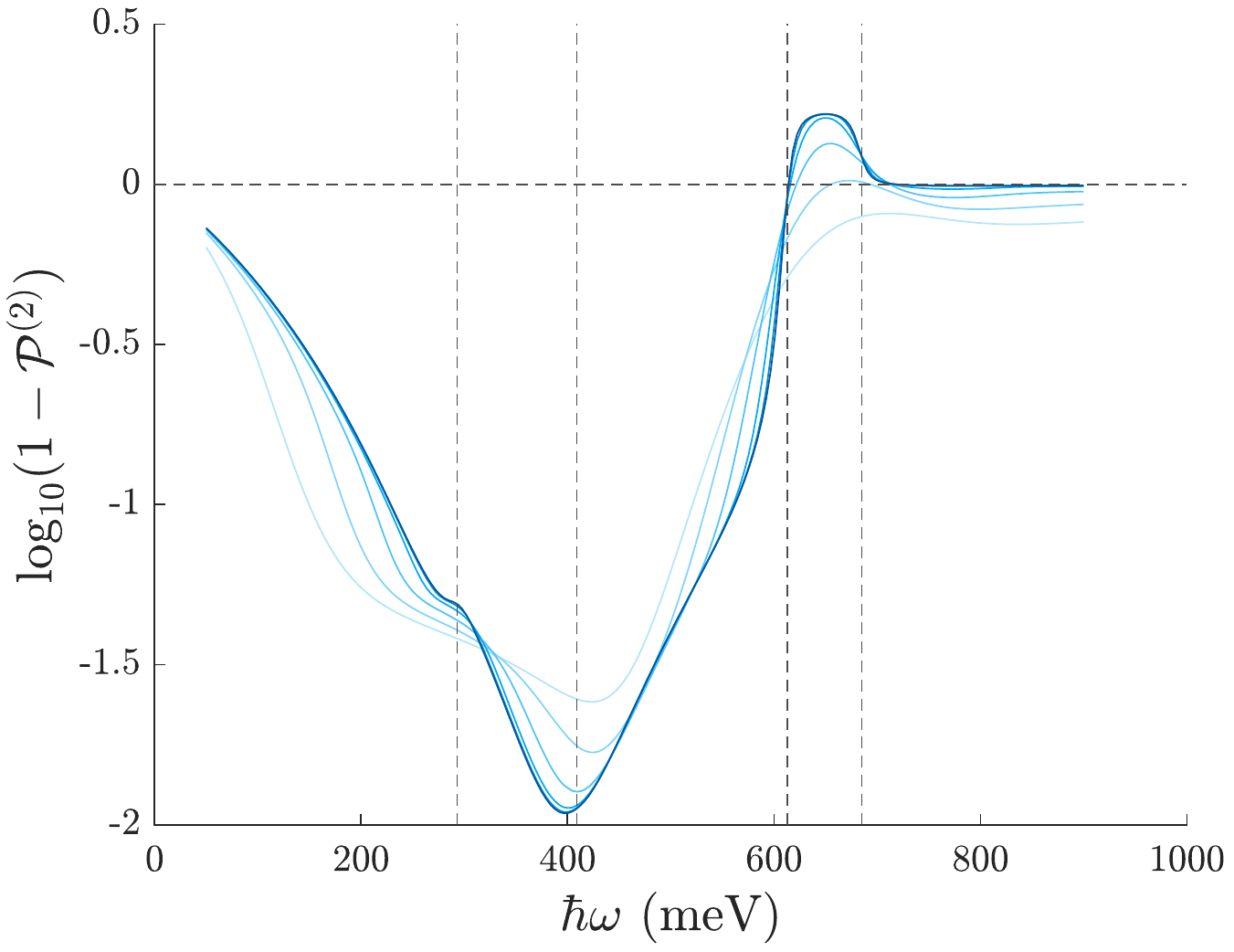}%
        }
    \subfloat[\label{effects:varytau0}]{%
    \includegraphics[width=0.5\linewidth]{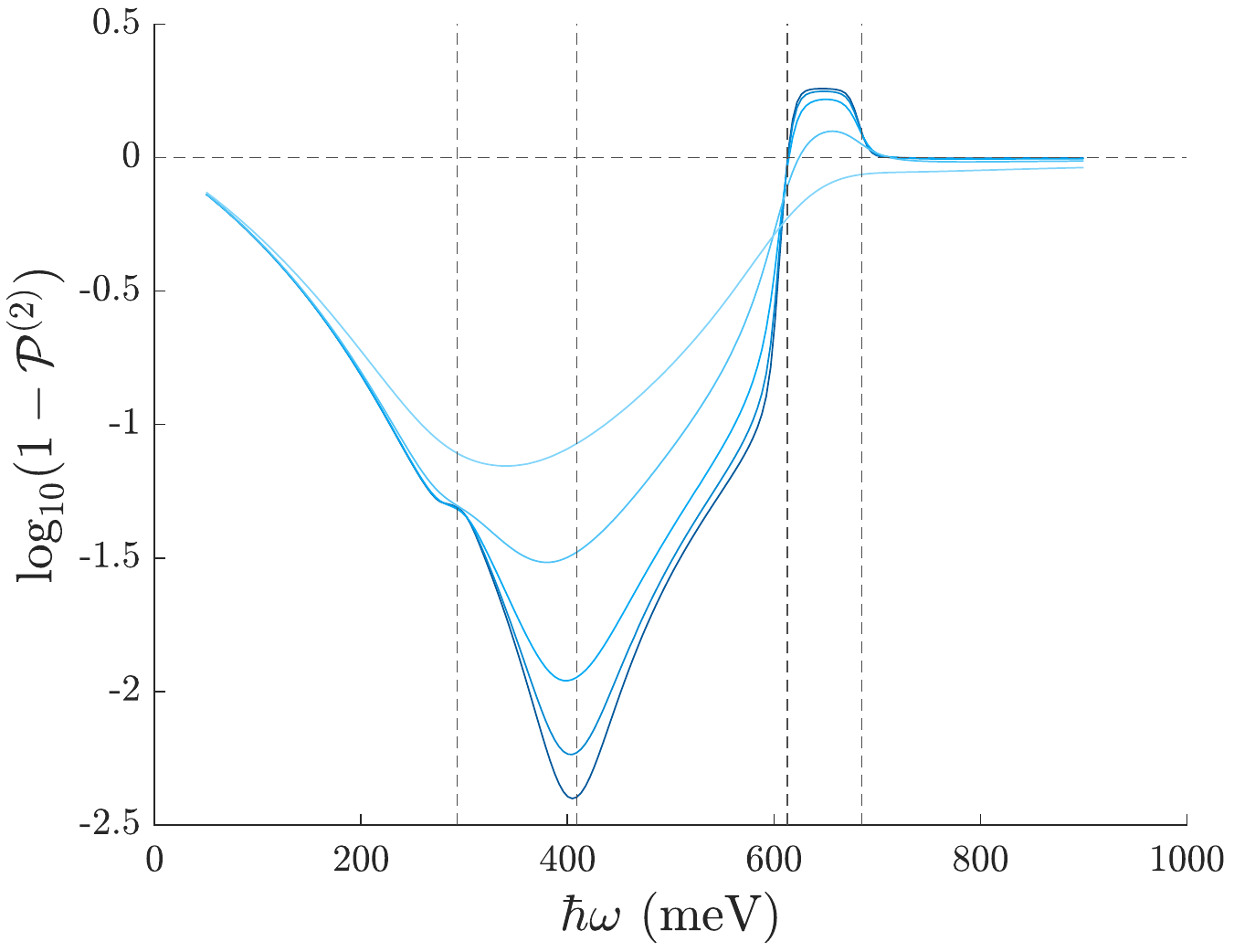}%
        }
        \caption{Deviation of the second-order valley-polarization from perfect polarization $\text{log}_{10}(1-\mathcal{P}^{(2)})$ as a function of the central photon energy $\hbar\omega$ of the exciting Gaussian pulse for a fixed bias of $a=205$ meV. In order of increasing energy, the vertical lines indicate the band gap energy, $\hbar\omega=2a,$ the onset of intervalley scattering, and the onset of down-scattering. The horizontal line indicates $\mathcal{P}^{(2)}=0.$ (a) From lightest to darkest, $t_p=10,15,25,50,100,$ and 200 fs for a fixed decoherence time of $\tau_0=100$ fs. The valley-polarization does not visibly change for $t_p>\tau_0.$ (b) From lightest to darkest, $\tau_0=10,30,100,200,$ and 300 fs for a fixed pulse duration of $t_p=100$ fs. Note that the scale of the valley-polarization axis is different in (a) and (b). In both (a) and (b), the valley-polarization is evaluated at $t_f=5t_p,$ the temperature is $300$ K, and the chemical potential is $\mu=0.$}\label{effects}
\end{figure*}

In this section, we study the effects of varying the pulse duration $t_p$ and decoherence time $\tau_0$ on the second-order valley-polarization. As might be anticipated, the effect of increasing $t_p$ and $\tau_0$ is to improve the valley-polarization and sharpen the response, as we explore in Fig. \ref{effects}. In Fig.~\subref*{effects:varytp} we fix the bias to a constant value of $a=205$~meV, and plot the second-order valley-polarization as a function of the central photon energy $\hbar\omega$ of the exciting pulse.
We use the scattering model developed in the previous section, accounting for both intervalley scattering (IVS) and down-scattering (DS). Thus, Fig.~\subref*{effects:varytp} can be thought of as a vertical slice of Fig.~\subref*{scattering:DS} along constant $a.$ In general, the valley-polarization increases steadily as $\hbar\omega$ approaches $2a,$ stalling briefly at the band gap energy, for reasons discussed in Sec. \ref{NoScattering}. The valley-polarization peaks close to $\hbar\omega=2a,$ then decays. The decay is accelerated by the presence of IVS, and an inverse valley-polarization is observed. The presence of DS counteracts the decay due to IVS. These general features are observed for all biases, but are particularly prominent for $a=205$ meV. In Fig.~\subref*{effects:varytp}, the band gap energy, $\hbar\omega=2a,$ and onsets of IVS and DS can be clearly seen (indicated with dashed vertical lines). 

Six different pulse durations are considered in Fig.~\subref*{effects:varytp}, and the response is evaluated at $t_f=5t_p$ for a relatively long decoherence time of $\tau_0=100$ fs \footnote{We note that in Fig.~\protect\subref*{effects:varytp} we have used pulse durations which can be shorter than a single cycle for some frequencies. While this may be unphysical or difficult to achieve experimentally, it is useful for demonstrating the effects of varying the pulse duration.}. As can be seen, increasing the pulse duration tends to sharpen the response about $\hbar\omega=2a$ and improve the maximum valley-polarization. For a $t_p=10$ fs pulse, a maximum valley-polarization of $\mathcal{P}^{(2)}=0.976$ is achieved, while for a $t_p=100$ fs pulse, the maximum valley-polarization achieved is $\mathcal{P}^{(2)}=0.989.$ Note that the $t_p=100$ fs curve and the $t_p=200$ fs curve are virtually indistinguishable. This can be attributed to the finite decoherence time (which results in linewidth broadening), making targeting $E_c(k)=a$ precisely impossible even when the pulse duration is very long. Therefore, from the point of view of valley-polarization, there is no reason to use a pulse duration that is much longer than the decoherence time. On the other hand, from the perspective of carrier density, using a very long pulse duration can be beneficial.

In Fig.~\subref*{effects:varytp}, the optimal operating frequency does not in general coincide exactly with $\hbar\omega=2a,$ and in fact varies somewhat with the pulse duration. For a $t_p=10$~fs pulse, the optimal operating frequency is $\hbar\omega=422$ meV, while for a $t_p=100$ fs pulse, the optimal operating frequency is $\hbar\omega=400$ meV. The value of the optimal operating frequency is a result of a complex interplay between the pulse duration, decoherence time, external bias (with corresponding $A/B$ ratio and density of states), the presence of intervalley scattering, and perhaps other factors. Taken together, these factors lead to violations of the simple rule of thumb that the optima lie along $\hbar\omega=2a.$ For the particular bias chosen in Fig.~\subref*{effects:varytp}, the presence of IVS seems to play an important role in the interplay, pushing the optima towards lower values as the pulse duration is increased. Some experimental fine-tuning may therefore be useful for finding the optimal operating frequency-bias pair for each particular sample and laser. As was discussed in Sec. \ref{NoScattering}, in the ideal limit of long pulse durations and decoherence times, the optimal operating frequency converges to $\hbar\omega=2a.$

In Fig.~\subref*{effects:varytau0} we again set the bias to $a=205$ meV, but this time fix the pulse duration to $t_p=100$ fs and vary the decoherence time. The same general features of Fig.~\subref*{effects:varytp} are observed. The main difference is that the maximum valley-polarization obtained in Fig.~\subref*{effects:varytau0} can be much larger than in Fig.~\subref*{effects:varytp} (note the different scales on the vertical axes). For a decoherence time of $\tau_0=10$~fs, a maximum valley-polarization of $\mathcal{P}^{(2)}=0.930$ is achieved, while for $\tau_0=300$ fs, the maximum valley-polarization achieved is $\mathcal{P}^{(2)}=0.996.$ The maximum valley-polarization continues to grow even as the decoherence time is increased through 500 fs (not shown). In Fig.~\subref*{effects:varytp}, improving the valley-polarization by increasing the pulse duration was limited by the decoherence time. Conversely, in Fig.~\subref*{effects:varytau0}, the valley-polarization can be increased indefinitely by using very pure samples with long decoherence times, regardless of the pulse duration. Comparing Figs.~\subref*{effects:varytp} and \subref{effects:varytau0}, it is clear that the decoherence time is the limiting factor with respect to the valley-polarization. While clean samples are preferable, a very strong valley-polarization can still be achieved when the decoherence time is short.

\subsection{Thermal carriers}\label{Density}
Up to this point we have only considered the valley-polarization arising from the second-order response $\mathcal{P}^{(2)},$ and have so far neglected the background of thermal carriers. In this section, we examine how thermal carriers act to reduce the valley-polarization and comment on strategies for minimizing their impact. To account for the thermal background, we simply add the thermal carrier density to the injected carrier density [see Eq. (\ref{nvgen})] and compute the valley-polarization. That is, we write $\widetilde{n}_V=\widetilde{n}_V^{(2)}+n_V^{(0)},$ where $\widetilde{n}_V^{(2)}$ was defined in Sec. \ref{IntrabandScattering} to account for intraband scattering via optical phonons (both IVS and DS). Since the thermal carrier density is the same in both the $K$ and $K'$ valleys, this corresponds mathematically to adding a factor of $2n_K^{(0)}=2n_{K'}^{(0)}$ to the denominator of Eq. (\ref{Pmod}). The valley-polarization is now
\begin{equation}
    \mathcal{P} = \frac{\widetilde{n}^{(2)}_K-\widetilde{n}^{(2)}_{K'}}{\widetilde{n}^{(2)}_K+\widetilde{n}^{(2)}_{K'}+2n_K^{(0)}}.
\end{equation}
Since ${\widetilde{n}_V^{(2)}}\,{\propto}\,{E_0^2}$ [Eq.~(\ref{nsubV})], the field strength had no effect on the second-order valley-polarization, but it will become important now as the zeroth-order response does not depend on $E_0$. We take $E_0=1.5\times(\omega/\omega_0)^{1/2}$~\si{\kilo\volt\per\centi\metre}, with $\hbar\omega_0=50$ meV. By scaling the field with the pulse frequency, we ensure that the number of photons per pulse is fixed across all frequencies. The magnitude of $E_0$ is chosen to ensure that $\rho_{cc}^{(2)}(\mbfk)$ is at most $1\%$ for any $\mbfk,$ such that first-order perturbation theory is sufficient.

\begin{figure*}
\centering
\subfloat[$T=300$ K\label{evo:300}]{
  \includegraphics[width=0.47\linewidth]{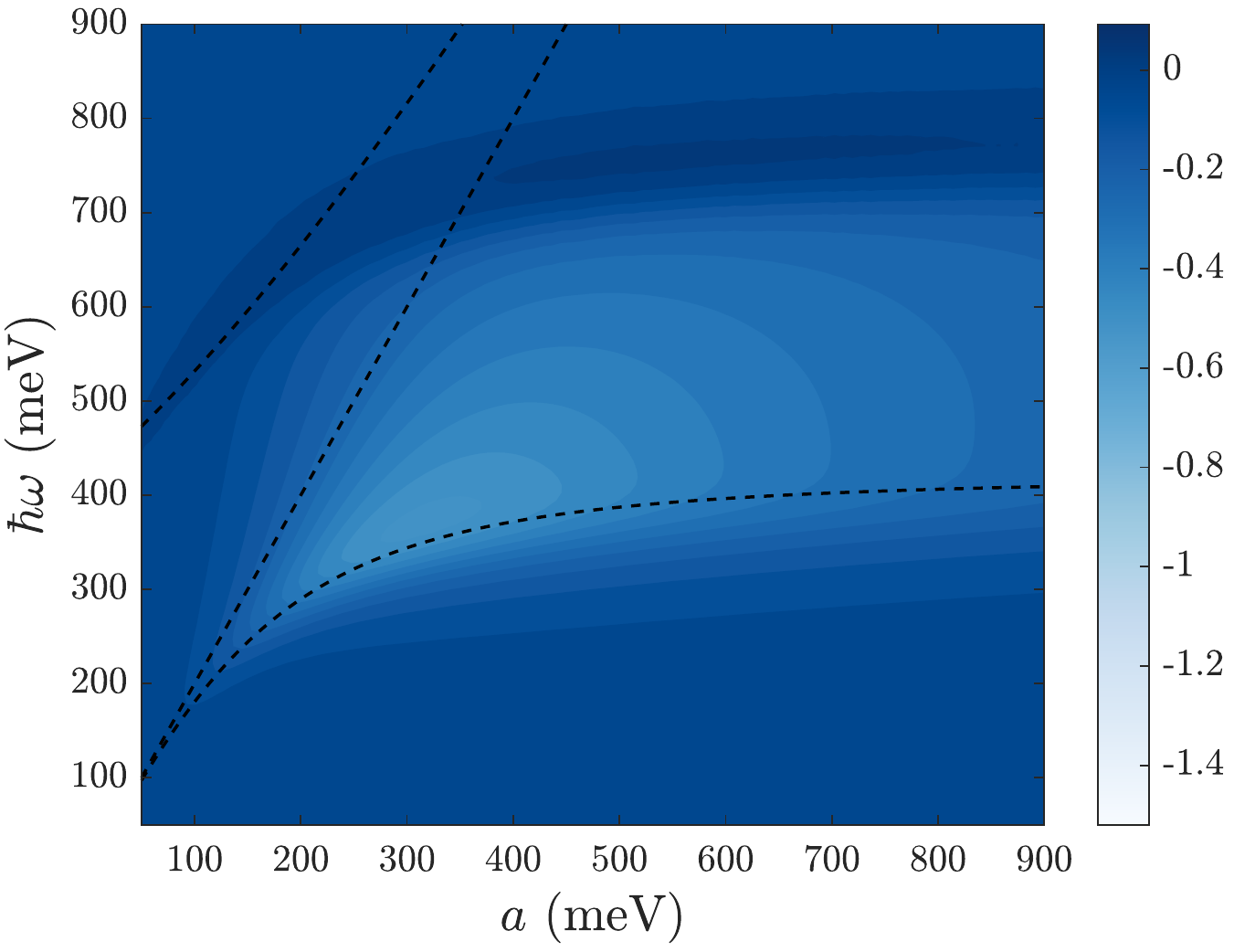}%
}
\subfloat[$T=250$ K\label{evo:250}]{%
  \includegraphics[width=0.47\linewidth]{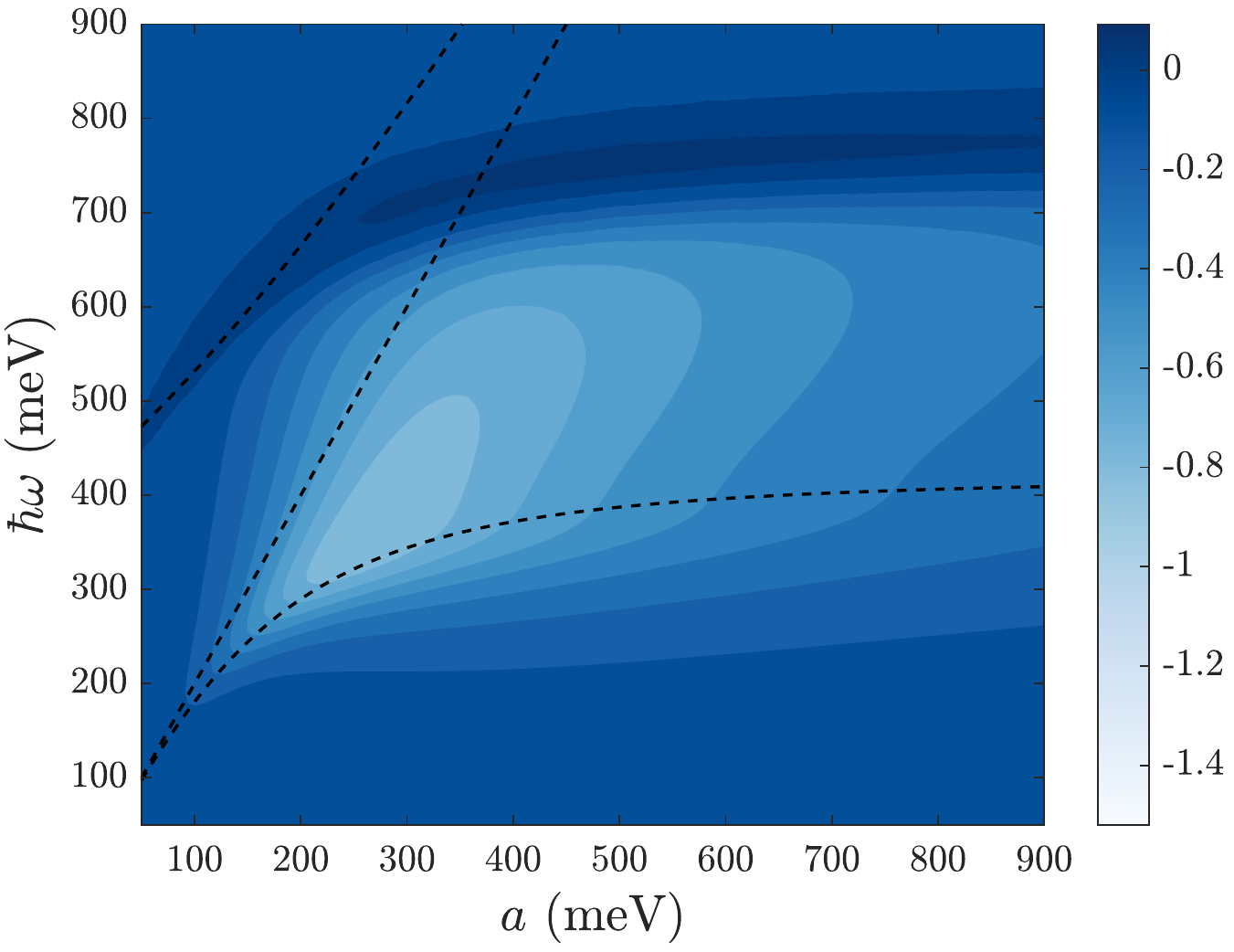}%
}

\subfloat[$T=200$ K\label{evo:200}]{%
  \includegraphics[width=0.47\linewidth]{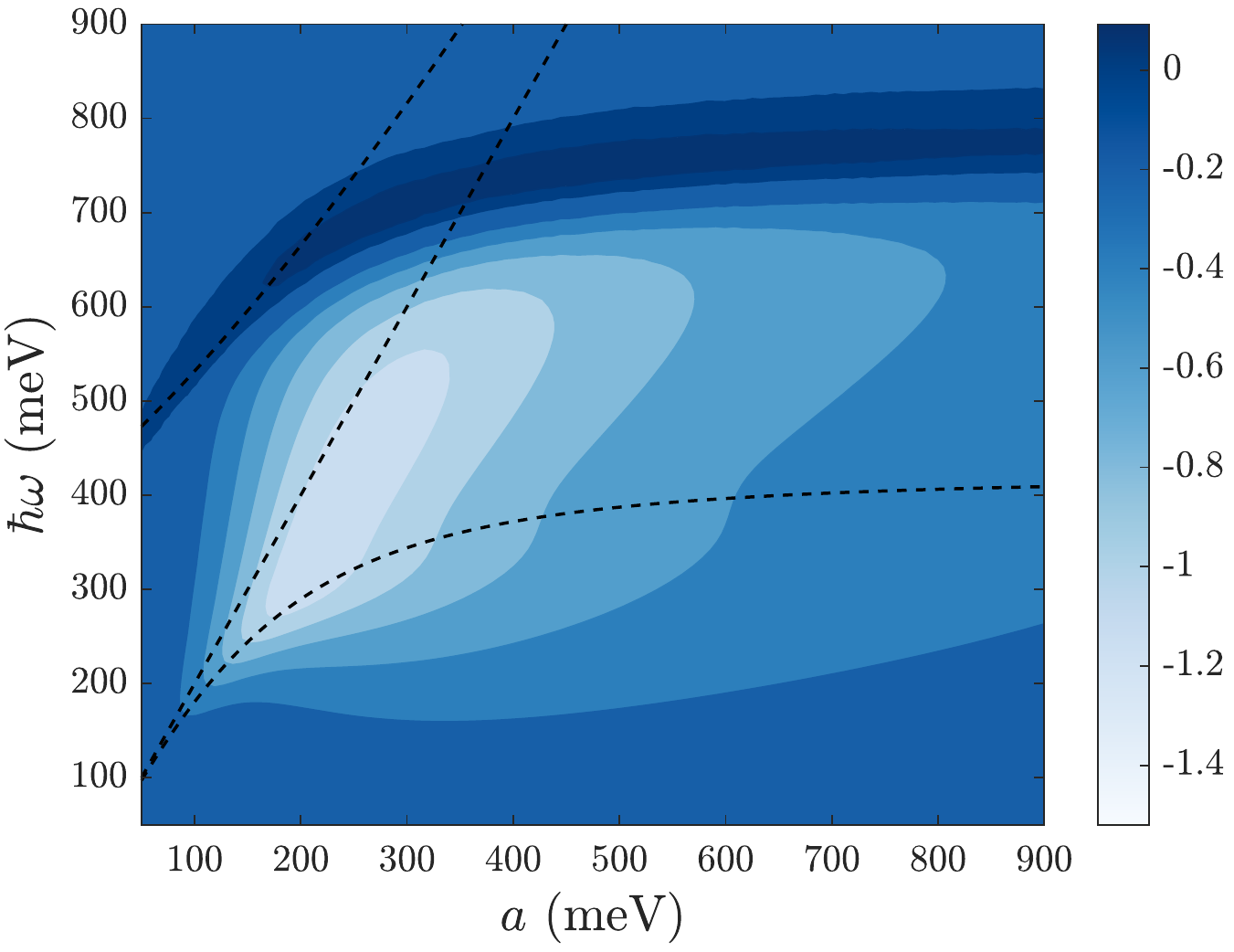}%
}
\subfloat[$T=150$ K\label{evo:150}]{%
  \includegraphics[width=0.47\linewidth]{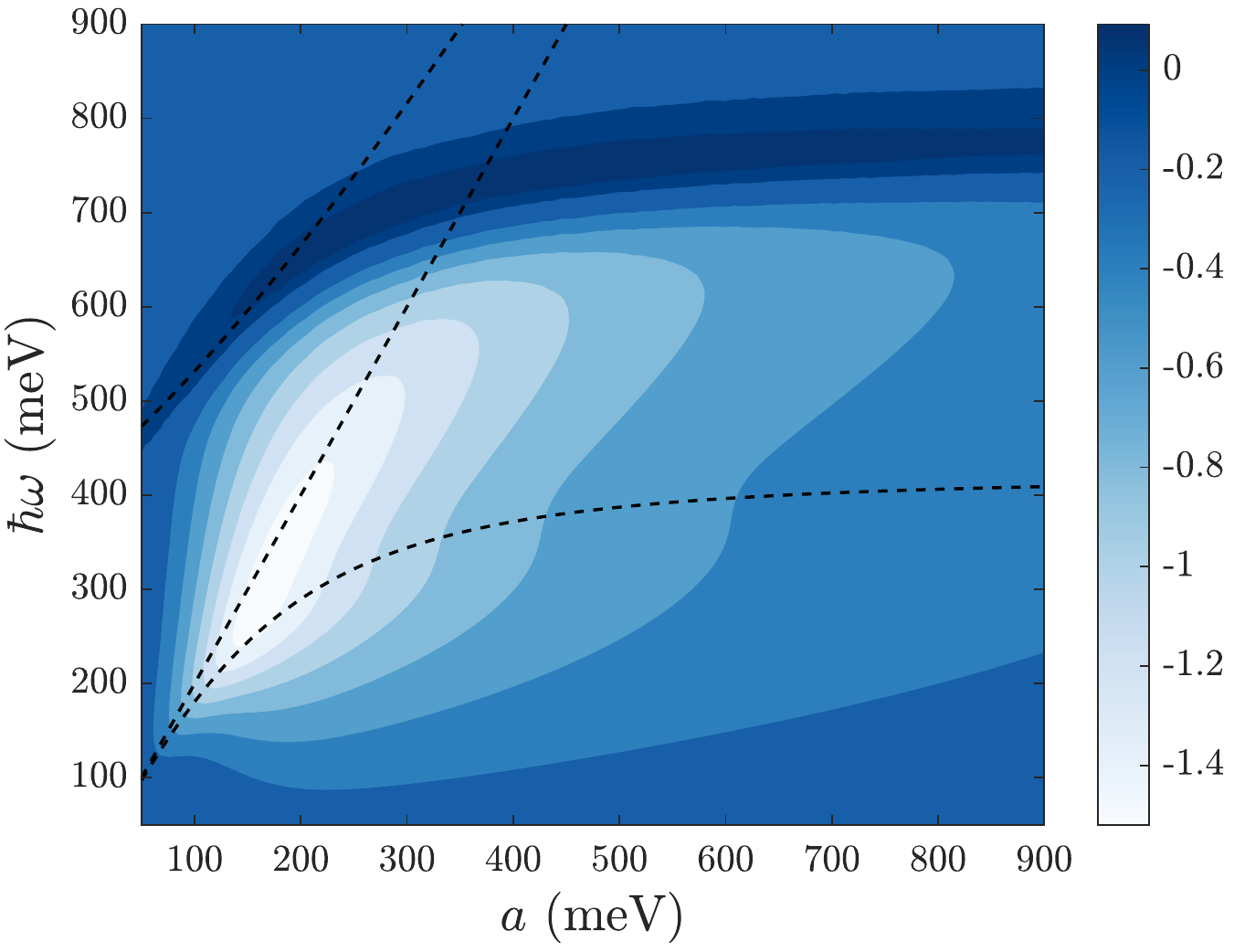}%
}
\caption{(Color online) Deviation of the valley-polarization (including thermal carriers) from perfect polarization $\text{log}_{10}(1-\mathcal{P})$ as a function of the external bias $a$ and central photon energy $\hbar\omega$ of the exciting Gaussian pulse at four different temperatures and including both intervalley scattering and down-scattering. The black dashed lines are as in Fig. \ref{VPNS}. The white dashed lines from Fig. \ref{scattering} have been omitted for clarity. The field amplitude is given by $E_0=1.5\times(\omega/\omega_0)^{1/2}$ \si{\kilo\volt\per\centi\metre}, with $\hbar\omega_0=50$ meV. The contour values are not the same for all four plots, but the color scale is consistent throughout. In all plots, the interband decoherence time is taken to be $\tau_0=30$ fs, the pulse duration is $t_p=50$ fs, and the chemical potential is $\mu=0.$} \label{evo}
\end{figure*}

In Fig.~\subref*{evo:300}, we plot the valley-polarization under the same conditions as Fig.~\subref*{scattering:DS} ($t_p=50$ fs, $\tau_0=30$ fs, $t_f=250$ fs, $T=300$ K, $\mu=0$), but this time account for the thermal background.
As can be seen, the valley-polarization is reduced significantly, with a maximum valley-polarization of $\mathcal{P}=0.70$ obtained at $(\hbar\omega,a)\approx(368,308)$ meV. The region of optimal operating frequency-bias pairs has also moved away from $\hbar\omega=2a$ and now hugs the band edge. This can be understood by considering the following. Along $\hbar\omega=2a,$ the second-order response is very strongly valley-polarized, however the actual number of carriers excited is quite small because we are operating away from the band edge where the joint density of states is small. Along the band edge, many more carriers are excited and even though the second-order response is not as pure, the \textit{difference} in carrier density between the $K$ and $K'$ valleys is able to overcome the thermal background. Note also that in contrast to Fig.~\subref*{scattering:DS}, the valley-polarization obtained below band gap is now essentially zero. This is because at frequencies below the band gap, very few carriers are excited and so the thermal population dominates.

In Figs.~\subref*{evo:300} through \subref{evo:150}, we show the evolution of the valley-polarization as the temperature is decreased from $300$ to $150$ K. As the temperature is decreased, the thermal carrier density decreases, so the valley-polarization improves and the optimal operating region returns to $\hbar\omega=2a.$ At $150$ K, the optimal operating frequency-bias pair is $(\hbar\omega,a)\approx(235,126)$ meV, yielding a very strong valley-polarization of $\mathcal{P}^{(2)}=0.97.$ However, even at $150$ K, thermal carriers significantly reduce the valley-polarization for small biases. This is because the band gap $\Delta E(a)\approx 2a$ for small biases, and so the thermal population is non-negligible. The presence of thermal carriers effectively cuts off small biases from the optimal operating region, much in the same way as intervalley scattering cut off high frequencies (Sec. \ref{IntrabandScattering}). Small biases may be desirable due to reduced electron mobility with increasing external bias $a$ \cite{borysDOS}. For $T=50$ K (not shown), the thermal population is negligible, and the valley-polarization observed in Fig.~\subref*{scattering:DS} is restored. In order to operate in the low-bias regime, one must work at low temperatures. If one works at a larger bias, a very strong valley-polarization can be achieved at 150 K.

It would be nice if a strong valley-polarization could be achieved at room temperature. One possibility is to increase the field strength, which has the effect of mitigating the thermal population, and follows a similar progression as Fig. \ref{evo}. We must be careful with increasing the field strength however, because as alluded to earlier, we begin to push the limits of first-order perturbation theory. To restore Fig.~\subref*{evo:300} to Fig.~\subref*{scattering:DS} by increasing the field amplitude alone, $E_0$ would need to be increased by about a factor of 50. Since $\rho_{cc}^{(2)}(\mbfk)\propto E_0^2,$ this puts us well outside the realm of first-order perturbation theory \cite{al-naib_optimizing_2015,luke,yaononlinear1,yaononlinear2}.

\section{Conclusion}\label{Conclusion}
In this work, we present a detailed study of valley-polarization in biased bilayer graphene. The energy bands are calculated using a tight-binding model, and density matrix equations of motion are derived within the length gauge. Electron populations in the $K$ and $K'$ valleys are calculated for an incident circularly polarized Gaussian pulse. The resulting population imbalance between the valleys is quantified by the valley-polarization, which we seek to maximize with respect to the external bias, pulse frequency, and pulse duration.

In the ideal limit (omitting thermal carriers, taking the pulse duration to be infinite, and neglecting carrier scattering and decoherence), we find that a perfect valley-polarization can be achieved when operating at pulse frequencies $\omega$ satisfying $\hbar\omega=2a,$ where $2a$ is the potential energy difference (the external bias) between the graphene layers. This result originates from a $k$-dependent valley-contrasting optical selection rule that becomes exact when $\hbar\omega=2a.$ In the presence of interband decoherence or finite pulse durations, it is not possible to achieve a perfect valley-polarization, even at zero temperature. However, a near-perfect valley-polarization ($>98\%$) can still be achieved by operating close to $\hbar\omega=2a.$

Intervalley scattering and thermal electron populations complicate the simple picture that the optimal operating frequency-bias pairs lie along the line $\hbar\omega=2a.$ While intervalley scattering via optical phonons has little effect on the valley-polarization for low-frequency pulses, we find that intervalley scattering greatly reduces the valley-polarization when the central photon frequency is large enough to inject carriers to an energy greater than that of a $K$-optical phonon. When we account for thermal populations, we find that the valley-polarization is significantly reduced when the external bias (and hence the band gap) is small. Thermal populations drive the optimal operating frequency-bias pairs away from $\hbar\omega=2a$ and towards the band edge, where the density of states is greatest. At room temperature, thermal populations reduce the maximum obtainable valley-polarization to just $70\%.$ However, the effect of thermal populations can be mitigated by working at low temperatures.

While the valley-polarization can be improved significantly by limiting defects and impurities, a strong valley-polarization can be achieved in any reasonably pure sample. To maximize the valley-polarization, the pulse duration should be close to or larger than the interband decoherence time of the sample. For a decoherence time of $30$ femtoseconds and a pulse duration of $50$ fs, we obtain the following optimal operating conditions. For room-temperature experiments, the optimal operating frequency-bias pair is $(\hbar\omega,a)=(368,308)$ meV, for which a valley-polarization of $70\%$ is obtained. At low temperatures ($T<150$ K), the optimal condition is $(\hbar\omega,a)=(235,126)$ meV, where a valley-polarization of over $97\%$ can be achieved. Given these promising results, we believe that bilayer graphene is a strong candidate for valleytronic applications.

\begin{acknowledgments}
This work was supported by Queen’s University and the Natural Sciences and Engineering Research Council of Canada (NSERC).
\end{acknowledgments}

\appendix
\section{Elliptical polarization}\label{Elliptical}
In this section, we examine why circularly polarized light is optimal to induce a valley-polarization in biased bilayer graphene. We also explore whether a general elliptical polarization is ever preferable to circular polarization. Consider a general electric field
\begin{equation}\label{generalfield}
    \mathbf{E}(t)=\alpha\mathbf{E}_L + \beta\mathbf{E}_R +c.c,
\end{equation}
which we have written in the basis of left- and right-hand circularly polarized components
\begin{align}
    \mathbf{E}_L&=(\mathbf{\hat{k}}- i\boldsymbol{\hat{\theta}}_k)e^{-i(\omega t +\theta_k)},\\
    \mathbf{E}_R&=(\mathbf{\hat{k}}+ i\boldsymbol{\hat{\theta}}_k)e^{-i(\omega t -\theta_k)},
\end{align}
expressed in polar coordinates with origin at either $\mathbf{K}$ or $\mathbf{K}'$ (the field takes the same form in both valleys). Here, $\alpha$ and $\beta$ are complex coefficients that characterize the polarization of $\mathbf{E}(t),$ with $|\alpha|^2+|\beta|^2=1.$ If $\alpha=0$ or $\beta=0$, $\mathbf{E}(t)$ is circularly polarized. If $|\alpha|=|\beta|,$ $\mathbf{E}(t)$ is linearly polarized, the polarization axis determined by the phase difference. If $\alpha\neq\beta$ and neither $\alpha$ nor $\beta$ are zero, we have some general elliptical polarization.

The first-order interband coherence $\rho_{cv}^{(1)}(\mbfk)$ is proportional to the carrier-field interaction $\boldsymbol{\xi}_{cv}(\mbfk)\cdot\mathbf{E}(t)$ [see Eq. (\ref{rhocv})].  If $\boldsymbol{\xi}_{cv}(\mbfk)\cdot\mathbf{E}(t)$ can be forced to zero in one valley but not the other, a strong valley-polarization is expected. For the field of Eq. (\ref{generalfield}), the carrier-field interaction takes the form
\begin{align*}\label{resint}
    \boldsymbol{\xi}_{cv}(\mathbf{k})\cdot\mathbf{E}(t) &\approx ie^{-i\omega t}
    \left[A(\alpha e^{-i\theta_k}+\beta e^{i\theta_k} )\right.\\&\left.\qquad\qquad\quad\pm B(-\alpha e^{-i\theta_k}+\beta e^{i\theta_k} )\right], \numberthis{}
\end{align*}
where the approximation sign indicates that we are considering only the resonant contributions, and where the plus and minus signs correspond to the $K$ and $K'$ valleys respectively. Here, $A$ and $B$ are the real, positive, $k$-dependent functions which were introduced in Sec. \ref{Connection}. Setting Eq. (\ref{resint}) to zero, we obtain the condition
\begin{equation}\label{alphabeta}
    \beta(B\pm A)=\alpha(B\mp A)e^{-2i\theta_k},
\end{equation}
which is valid for any choice of gauge. The problem of forcing the resonant piece of $\boldsymbol{\xi}_{cv}(\mathbf{k})\cdot\mathbf{E}(t)=0$ has been reduced to satisfying Eq. (\ref{alphabeta}). On the surface, we are faced with serious problem: Eq. (\ref{alphabeta}) depends on $\theta_k,$ suggesting that Eq. (\ref{alphabeta}) can only be satisfied along a single radial direction in $k$-space. Since $A$ and $B$ are $k$-dependent, this then implies that Eq. (\ref{alphabeta}) can only be satisfied at (at most) a couple of \textit{points} in $k$-space. One way to circumvent this issue is to use circularly polarized light. If we set $\alpha=0$ (right-hand circular polarization) we obtain
\begin{equation}
    (B\pm A)=0,
\end{equation}
which can be satisfied in the $K'$ valley when $A=B,$ but never in $K$ \footnote{That is, unless $A=B=0,$ but then no carriers are injected in either valley}. Similarly, if we set $\beta=0$ (left-hand circular polarization), we obtain 
\begin{equation}
    0=(B\mp A),
\end{equation}
which can be satisfied in the $K$ valley when $A=B,$ but never in $K'.$ Thus, for any non-circular elliptical polarization, the phase-dependence of Eq. (\ref{alphabeta}) limits the optical selection rule to just a couple of $k$-space points. By choosing circularly polarized light, the phase dependence of Eq. (\ref{alphabeta}) is removed, and a valley-contrasting optical selection rule is obtained for the specific constraint $A=B.$ Since $A=B$ occurs for $k=k_1$ (see Sec. \ref{NoScattering}), the valley-contrasting optical selection rule is simultaneously satisfied around an entire ring of states in $k$-space, as opposed to just a couple of points. Circular polarization is therefore always preferable to elliptical.

\section{Complex-shifted integral}\label{CSI}
In Eq. (\ref{postcompl}), $\lambda$ is in general complex, but the integral bounded by $(-\infty,\infty)$ is equivalent to a complex-shifted integral bounded by $(-\infty+i\gamma,\infty+i\gamma)$ for some real $\gamma$.

\textit{Proof:} By Cauchy's integral theorem, an integral over the closed contour $C = [-R,R,R+i\gamma,-R+i\gamma,-R]$ is zero:
\begin{align*}
    0 = {\oint_C} {f(\eta)} d\eta\, {=} {\int_{-R}^{R}} {+} {\int_{R}^{R+i\gamma}} {+} {\int_{R+i\gamma}^{-R+i\gamma}} {+} {\int_{-R+i\gamma}^{-R}}{,} \numberthis{}
\end{align*}
where $R$ is real and ${f(\eta)=\left(\text{erf}\left(\eta\right)+1\right)\text{exp}\left(-(\eta+c)^2\right)}.$ For $R\rightarrow\infty$, the integrals over the segments of constant $R$ are zero because $f(\eta)$ goes to zero. Thus, the integral along the real axis is equal to an arbitrarily complex-shifted integral over the same real limits
\begin{align*}
    0= \int_{-R}^{R} + \int_{R+i\gamma}^{-R+i\gamma}= \int_{-R}^{R} - \int_{-R+i\gamma}^{R+i\gamma}.\numberthis{}
\end{align*}

\bibliographystyle{apsrev4-1}
\bibliography{biblio}

\biblio

\end{document}